\documentclass[reprint]{JASAnew}

\newcommand{\GTPE}{$\mathrm{WAPE}_{\mathrm{topo}}$}
\definecolor{mblue}{rgb}{0,0.447,0.741}

\definecolor{track}{rgb}{0,0,0}
\begin{document}

\title[Wind turbine sound propagation]{Wind turbine sound propagation: Comparison of a linearized Euler equations model with parabolic equation methods}

\author{Jules Colas}
\affiliation{Univ Lyon, Ecole Centrale de Lyon, CNRS, Univ Claude Bernard Lyon 1, INSA Lyon, LMFA, UMR5509, 69134 Ecully CEDEX, France}
\author{Ariane Emmanuelli}
\affiliation{Univ Lyon, Ecole Centrale de Lyon, CNRS, Univ Claude Bernard Lyon 1, INSA Lyon, LMFA, UMR5509, 69134 Ecully CEDEX, France}
\author{Didier Dragna}
\affiliation{Univ Lyon, Ecole Centrale de Lyon, CNRS, Univ Claude Bernard Lyon 1, INSA Lyon, LMFA, UMR5509, 69134 Ecully CEDEX, France}
\author{Phillipe Blanc-Benon}
\affiliation{Univ Lyon, CNRS, Ecole Centrale de Lyon, INSA Lyon, Univ Claude Bernard Lyon 1, LMFA, UMR5509, 69134 Ecully CEDEX, France}
\author{Benjamin Cotté}
\affiliation{Institute of Mechanical Sciences and Industrial Applications (IMSIA),  ENSTA Paris, CNRS, CEA, EDF, Institut Polytechnique de Paris}
\author{Richard Stevens}
\affiliation{Physics of Fluids Group, Max Planck Center Twente for Complex Fluid Dynamics, J. M. Burgers Center for Fluid Dynamics, University of Twente, P. O. Box 217, 7500 AE Enschede, The Netherlands}

\begin{abstract}
  {\color{track} 
  Noise generated by wind turbines is significantly impacted by its propagation in the atmosphere. 
  Hence, for annoyance issues an accurate prediction of sound propagation is critical to determine noise levels around wind turbines. 
  } 
  This study presents a method to predict wind turbine sound propagation based on linearized Euler equations. We compare this approach to the parabolic equation method, which is widely used since it captures the influence of atmospheric refraction, ground reflection, and sound scattering at a low computational cost. 
  Using the linearized Euler equations is more computationally demanding but \replaced{more accurate}{can reproduce more physical effects} as fewer assumptions are made. 
  An additional benefit of the linearized Euler equations is that they provide a time-domain solution.
  To compare both approaches, we simulate sound propagation in two distinct scenarios. In the first scenario, a wind turbine is situated on flat terrain; in the second, a turbine is situated on a hilltop. The results show that both methods provide similar noise predictions in the two scenarios.
  We find that while some differences in the propagation results are observed in the second case, the final predictions for a broadband extended source are similar between the two methods.  
\end{abstract}

\maketitle

\section{Introduction}

The noise produced by wind turbines is one of the main constraints for the installation of new wind farms. It can also entail curtailment plans when annoyance issues emerge after the construction of the farm leading to energy production loss up to 70$~\%$  during the night \citep{dumortier_acoustic_2015}. 
Therefore, accurate prediction tools are needed to assess the wind farm noise during the development and operational phases. 
The modeling of the aerodynamic noise sources and the propagation of sound in the atmosphere are two critical fields of study for the prediction of wind turbine noise. 
Several models for the noise emitted by a wind turbine have been developed in recent years \citep{cotte_extended_2019, barlas_consistent_2017}. 
These models aim at capturing both the mean sound pressure level (SPL) emitted by the source, and the amplitude modulation (AM) induced by the rotation of the blades. 
The unsteady nature of the source is considered to be one of the main annoyance causes as the global SPL is usually quite low \citep{hansen_prevalence_2019}. 
To predict the SPL field around the turbine, outdoor sound propagation models need to consider ground effects, atmospheric absorption, and refraction induced by the variability of wind and temperature within the atmospheric boundary layer (ABL). 
The effect of the ABL flow on wind turbine sound propagation has been extensively studied ~\cite{barlas_variability_2018,heimann_3d-simulation_2018}. 
The evolution of the temperature and wind speed gradients during the day modifies the sound propagation and, hence, the SPL around the wind turbine. 
Furthermore, the presence of the wind turbine creates a wake that acts as a waveguide and tends to increase the SPL \added{at a specific location} downwind of the turbine \citep{barlas_effects_2017}. 
Finally, topography was also shown to play a significant role in wind turbine sound  propagation~\citep{heimannSoundPropagationWake2018,sessarego_noise_2020,shen_advanced_2019}.  

Various numerical methods have been used to calculate wind turbine sound propagation. 
Engineering tools based on simple empirical models are more suited for operational purposes. Methods based on \replaced{geometrical}{ray} acoustics can consider atmospheric refraction \citep{heimann_3d-simulation_2018, prospathopoulos_application_2007}. 
However, they are usually less precise in terms of SPL and rely on a high-frequency approximation. The parabolic equation (PE) methods have been used extensively both for their good accuracy at long range and for their low computational cost \citep{gilbert_application_1989}. 
They consist in solving a one-way-wave equation in the frequency domain usually in a two-dimensional (2D) geometry, although the formulation holds in three dimensions (3D). PE methods have thus become the state of the art for wind turbine noise propagation~\citep{barlas_effects_2017,kayser_environmental_2020}. Nevertheless, they suffer from several limitations. First, the solution is only valid for propagation angles close to the main propagation direction. Second, the one-way wave equation neglects back-scattering. This could lead to inaccuracies in the presence of topography with steep slopes.
\replaced{Finally, errors can occur when taking into consideration a moving atmosphere.Indeed, the most common approach is to consider an inhomogeneous atmosphere at rest with an effective sound speed, that includes the wind velocity component along the propagation direction. This approach can be inaccurate if the wind direction is not aligned with the propagation direction. Improved PE methods consider the effect of the atmospheric boundary layer on sound propagation by including the mean flow terms when deriving the one-way wave equation \citep{dallois_wide-angle_2001,ostashev_wave_2020}.}{
Finally, approximations are usually made in order to consider propagation in an inhomogeneous moving medium.
The most common approach is to consider an atmosphere at rest with an effective sound speed, that includes the effect of both temperature and of the wind velocity gradient.
This approach can be inaccurate if the wind speed is too high or if its vertical component is not negligible \cite{dallois_wide-angle_2001}. 
Improved PE methods consider the effect of the ABL on sound propagation by including the mean flow terms when deriving the one-way wave equation \citep{dallois_wide-angle_2001,ostashev_wave_2020}.
}
The range of application of PE methods for wind turbine noise has already been investigated in several studies, by comparing PE results to analytical solutions or measurements  \citep{lee_prediction_2016,nyborg_propagation_2022,kayserValidityEffectiveSound2023}. However, validation against analytical solutions is restricted to \replaced{academic}{simple} cases.
In addition, outdoor experiments are not perfectly controlled, as, for instance, the wind field is \replaced{}{only} known partially or the wind turbine noise sources have to be modeled. Comparison against measurements remains thus overall limited. 

A more advanced sound propagation method is to solve the linearized Euler equations (LEE) directly. This method is also widely used for outdoor propagation \citep{salomons_eulerian_2002,blumrich_linearized_2002,dragna_towards_2014, van_renterghem_efficient_2014}. It takes the effect of the mean flow on sound propagation accurately into account. 
In particular, it considers the vertical component of the wind speed, which is neglected in the effective sound speed approach.
The LEE method overcomes the PE limitations, which include restricted angular validity and restrictions on backscattering.
In addition, as a time-domain method, it provides a broadband solution and can  be used to consider unsteady effects. 
The main drawback is the higher computational cost of the LEE compared to the PE method, which explains why it has not yet been considered for wind turbine noise propagation. 

The objectives of this paper are to introduce a LEE model for wind turbine noise propagation and to assess the advantages of using a LEE model with respect to state-of-the-art PE methods. 
The LEE model can account for topography, ground impedance, and inhomogeneous mean flow. 
For comparison, two different PE implementations are  considered: a vector PE able to handle strong wind variation but limited to flat terrain, and a PE formulation able to consider topography and wind gradients through an effective sound speed approach. 
We compare the results of the LEE and PE approaches for two realistic cases:  first a wind turbine on flat ground and second \replaced{}{a wind turbine located on top of a steep hill}. 
For both cases, the mean flow in which the sound propagates is obtained from previously computed large eddy simulations (LES) by \citet{liu_effects_2020} and an extended moving source model based on Amiet's theory~\citep{cotte_extended_2019} is used.

The paper is organized as follows. In Sec.~\ref{sec:method} the complete methodology for the wind turbine noise prediction is described including the LEE and PE methods.  Then Sec.~\ref{sec:cases} details the two cases studied. In Sec.~\ref{sec:results} the SPL and AM obtained with both methods are compared for the two cases. Finally, concluding remarks are given in Sec.~\ref{sec:conclusion}.

\section{Method}
\label{sec:method}
\subsection{General methodology}
\label{sec:general_layout}
\begin{figure}[ht]
\centering
\includegraphics[]{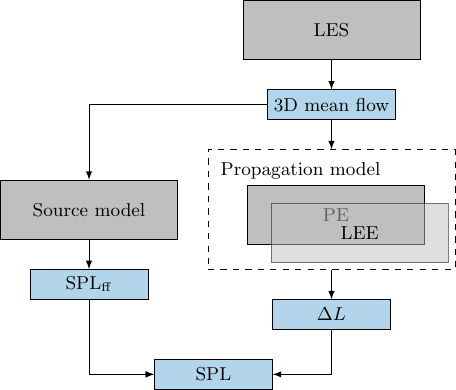}
\caption{Diagram of the complete prediction methodology}
\label{fig:framework}
\end{figure}

The general framework used to compute wind turbine noise at a receiver location is summarized in Fig.~\ref{fig:framework}.
It is based on the coupling of three different models: a LES code used to obtain the mean wind velocity in the ABL, a source model based on Amiet's theory, and a propagation model. 
The propagation models studied in this work are then detailed in Sec. \ref{sec:propa}.

First, the LES code is used to compute a realistic ABL.
The interaction between the flow and the wind turbine is modeled with an actuator disk method. 
This code has been extensively tested both for the computation of realistic atmospheric boundary layers \citep{gadde_large-eddy_2021} and for the simulation of wind farm flow, with or without topography \citep{gadde_interaction_2021,liu_effects_2020}. 
Although the simulations  are unsteady, only the mean velocity fields are used in the following. 
\replaced{}{By doing so the turbulence scattering is not considered in this study, although it is known to have an impact on the propagation of wind turbine noise \cite{barlas_effects_2017}.} 
The flow data are then fed into an extended source model and a propagation model based on either the LEE or on the PE.\@ 

The source model derives from Amiet's strip theory: Each blade is divided into several segments, considered as uncorrelated sources. 
For each segment the turbulent inflow and trailing edge noise are computed using the model developed by \citet{tian_wind_2016}.
The resulting SPL in free field, denoted SPL$_{\mathrm{ff}}$, depends on the wind turbine geometry (chord length and type of the blade segment) and on the incoming velocity profile and turbulent spectrum.
Additionally, to determine the propagation effects, such as ground reflection or atmospheric refraction, the SPL relative to the free field, denoted $\Delta L$, must be computed for each source-receiver pair. 

Finally, source and propagation effects are considered by computing the SPL at a receiver produced by one segment $i$ at a given angular position of the blades $\beta$ such that:
\begin{equation}
    \label{eq:salomons}
    \begin{aligned}
    \mathrm{SPL}^i(\mathbf{x},\omega,\beta) = & {\rm SPL}_{\rm ff }^i(\mathbf{x},\omega,\beta) \\
     & +\Delta L^i(\mathbf{x},\omega,\beta) -\alpha(\omega) R, 
    \end{aligned}
\end{equation}
where  $\mathbf{x}$ is the receiver coordinates,  $\omega$  the angular frequency,  $\alpha$ the atmospheric absorption coefficient\replaced{}{,} and $R$ the distance from source to receiver.
The contributions of each segment are combined to obtain the total SPL at the receiver:
\begin{equation}
    \begin{aligned}
        {\rm SPL}(\mathbf{x},\omega,\beta) = 10\log_{10}\left(\sum_{i=1}^{N_s} 10\,^{{\rm SPL}^i(\mathbf{x},\omega,\beta)/10}\right),
    \end{aligned}
\end{equation}
where $N_s$ is the number of segments used  to discretize the three blades. 
The overall SPL (OASPL) are then computed by summing the SPL over $\omega$ for a given angular position of the blades.
In addition, amplitude modulation can be quantified by measuring the difference between the minimum and maximum OASPL values during one rotation of the blades.
This method can become very expensive as the number of propagation simulations is proportional to the number of receivers, angular positions, and blade segments. 

\begin{figure}[ht]
\centering
\includegraphics[]{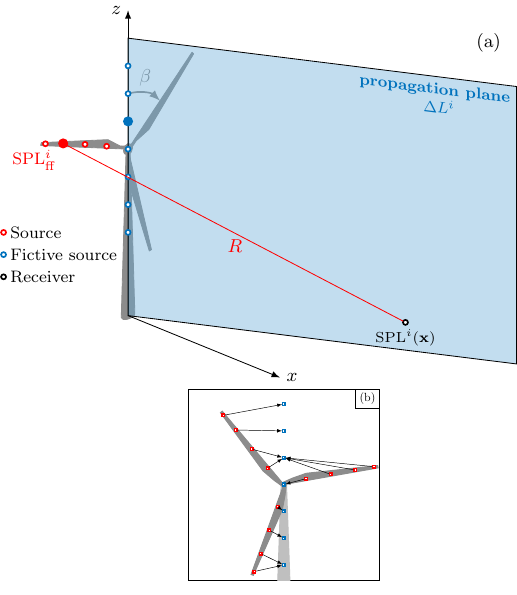}
\caption{(a) Calculation of the SPL at a receiver due to one blade segment. (b) Total SPL from all segments is determined using the closest fictive source to compute the $\Delta L$.}
\label{fig:coupling}
\end{figure}

To limit the computational cost, the method proposed by \citet{cotte_extended_2019} considers a number of fictive source heights distributed over a vertical line in the rotor plane, as depicted in Fig.~\ref{fig:coupling}\textcolor{mblue}{.a}.
The propagation simulations are performed only for these fictive sources. 
Subsequently, the  SPL$^{i}$ in Eq.~(\ref{eq:salomons}) is computed using the value of $\Delta L$ corresponding to the closest fictive source, see Fig.~\ref{fig:coupling}\textcolor{mblue}{b}. 
Note that the number of propagation simulations is reduced by considering a small number of source positions and by positioning all fictive sources in the same propagation plane. 
Hence, the same $\Delta L$ results are utilized to compute SPL for all receivers in this plane.
This method is based on the assumption that the dimension of the wind turbine rotor is relatively small in relation to the propagation distances.
\replaced{
To use this method it is also important to ensure that the number of fictive source heights is sufficient to obtain convergence of the results, \textit{i.e.} that the average SPL and AM during a rotation do not depend on the number of sources used. 
In \citet{cotte_extended_2019}, it was shown that a minimum of seven source heights is necessary to obtain this convergence.
}{
  In order to use this method it is also important to ensure that the number of fictive source heights is sufficient to obtain convergence of the results \textit{i.e.} that the average SPL and AM during a rotation do not depend on the number of sources used. 
  In \citet{cotte_extended_2019}, it was shown that a minimum of 7 source heights is necessary to obtain this convergence in the downwind direction.
  }

\subsection{Acoustic propagation simulation}
\label{sec:propa}
In this section the models used to predict the sound propagation in the ABL are introduced. 
First, the LEE model developed for this study is presented and then the PE implementations are described.

\subsubsection{LEE}
\paragraph{Acoustic model.}
This propagation model is based on the finite-difference solution of the LEE in a curvilinear mesh \citep{dragna_towards_2014}. 
Thus, the acoustic field is solved separately from the mean flow field and does not affect it. 
The LEE are derived from the conservation of mass, momentum, and energy for atmospheric propagation by neglecting terms of order $(|\mathbf{V_0}|/c_0)^2$ \citep{ostashev_equations_2005},
\begin{equation}\label{eq:LEE}
    \begin{aligned}
    &\frac{\partial p}{\partial t} + (\mathbf{V_0} \cdot \nabla) p +\rho_0 c_0^2 \nabla \cdot \mathbf{v}=0 \;,  \\
    &\frac{\partial \mathbf{v}}{\partial t} + (\mathbf{V_0} \cdot \nabla) \mathbf{v} +(\mathbf{v} \cdot \nabla) \mathbf{V}_{\mathbf{0}}+\frac{\nabla p}{\rho_0}=0 \;,
    \end{aligned}
\end{equation}
where \replaced{}{$t$ is the time}; $p$ and \replaced{$ \mathbf{v} = (u,v)$}{$ \mathbf{v} = (u,w)$} are  the acoustic pressure and velocity; and $\rho_0$, $c_0$, and \replaced{$\mathbf{V_0}~=~(u_0,v_0)$}{$\mathbf{V_0}~=~(u_0,w_0)$} are the mean density, sound speed, and velocity. 
The 2D LEE in the $(x,z)$ plane can be written in conservative form as 
\begin{equation}
    \mathbf{U}_t + \mathbf{E}_x +\mathbf{F}_z + \mathbf{H} = 0 \; ,
\label{eq:conservative}
\end{equation}
where \replaced{$\mathbf{U}=[p, \rho_0 u, \rho_0 v]^T$}{$\mathbf{U}=[p, \rho_0 u, \rho_0 w]^T$} is the vector of unknowns; the partial derivatives are denoted  $i_j = \partial i/\partial j$; and $\mathbf{E}$, $\mathbf{F}$, and $\mathbf{H}$ are the Eulerian fluxes \replaced{}{defined as}
\replaced{ 
\begin{equation}
    \begin{aligned}
    \mathbf{E} &=&\left(\begin{array}{c}
            u_{0} p+\rho_0 c_0^2 u \\
            u_{0} \rho_0 u+p \\
            u_{0} \rho_0 v
            \end{array}\right),\\
    \mathbf{F} &=&\left(\begin{array}{c}
        v_{0} p+\rho_0 c_0^2 v \\
        v_{0} \rho_0 u \\
        v_{0} \rho_0 v+p
        \end{array}\right), \\
    \mathbf{H} &=&\left(\begin{array}{c}
        -p\left(\nabla \cdot \mathbf{V}_{\mathbf{0}}\right) \\
        \rho_0(\mathbf{v} . \nabla) u_{0} \\
        \rho_0(\mathbf{v} . \nabla) v_{0}
        \end{array}\right).
    \end{aligned}
    \end{equation}
}{
  \begin{equation}
    \begin{aligned}
    \mathbf{E} &=\left[u_{0} p+\rho_0 c_0^2 u, u_{0} \rho_0 u+p, u_{0} \rho_0 w\right]^T,\\
    \mathbf{F} &=\left[
        w_{0} p+\rho_0 c_0^2 w,
        w_{0} \rho_0 u, 
        w_{0} \rho_0 w+p
        \right]^T,\\
    \mathbf{H} &=\left[
        -p\left(\nabla \cdot \mathbf{V}_{\mathbf{0}}\right),
        \rho_0(\mathbf{v} . \nabla) u_{0},
        \rho_0(\mathbf{v} . \nabla) v_{0}
        \right]^T.
    \end{aligned}
    \end{equation}
}
\replaced{
The acoustic source is a Gaussian pulse. It is introduced via the initial conditions of Eq.~(\ref{eq:conservative}),
\begin{equation}
  \mathbf{U}(t=0)=\left[S_0 \exp \left(-\frac{R^2}{B^2}\right), 0,0\right]^T, B^2= \frac{(3\Delta x)^2}{\log(2)}
\end{equation}
with $R$ the distance to the source, $S_0 = 1~$Pa the amplitude of the source, and $\Delta x$ the streamwise grid spacing. 
}{
The acoustic source is a pulse with a Gaussian spatial distribution, introduced via the initial conditions of Eq.~(\ref{eq:conservative}):
\begin{equation}
  \mathbf{U}(x,z,t=0)=\left[S_0 \exp \left(-\frac{R^2}{B^2}\right), 0,0\right]^T ,
  \label{e:source}
\end{equation}
where $R =\sqrt{x^2 + (z-z_S)^2}$, $B^2= (3\Delta x)^2 / \log(2)$,  $z_S$ is the height of the center of the pulse, $S_0 = 1~$Pa is the amplitude of the source, and $\Delta x$ is the streamwise grid spacing.
This source aims to represent a broadband monopole, with frequency content up to $f = 0.6c_0/B$; see appendix Eq.~(\ref{eq:f_content}).
}

\paragraph{Curvilinear transformation.}
To take into account a ground profile $h$, a transformation  of the coordinate system is applied from Cartesian coordinates $(x,z)$ to curvilinear coordinates $(\xi,\eta)$ such that
\begin{equation}
    \begin{aligned}
    &x(\xi, \eta)=\xi, \\
    &z(\xi, \eta)=h(\xi)+\frac{\eta}{z_{\max }}\left[z_{\max }-h(\xi)\right],
    \end{aligned}
    \label{eq:curvi-transform}
\end{equation}
where $h(x)$ is the terrain elevation and $z_{\max}$ is the maximum height of the domain. 
Hence, Eq.~(\ref{eq:conservative}) becomes: 
\begin{equation}
    \left(\frac{\mathbf{U}}{J}\right)_t+\left(\frac{\xi_x \mathbf{E}+\xi_z \mathbf{F}}{J}\right)_\xi+\left(\frac{\eta_x \mathbf{E}+\eta_z \mathbf{F}}{J}\right)_\eta+\frac{\mathbf{H}}{J}=0, 
\label{eq:final}
\end{equation}
where $J~=~|\xi_x \eta_z-\xi_z \eta_x|$ is the Jacobian of the transformation. 
This transformation was proposed by \citet{gal-chen_use_1975} to recover a flat top boundary. 

\paragraph{Numerical scheme}
The formulation in Eq.~(\ref{eq:final}) can be written as
\begin{equation}
    \mathbf{U}_t = \mathcal{F}(\mathbf{U}),
\end{equation}
where $\mathcal{F}$ is a function of $\mathbf{U}$ and its spatial derivatives. 
A fourth-order six-step Runge-Kutta (RK) algorithm \citep{berland_low-dissipation_2006} is used to integrate the solution from $\mathbf{U}(t_n)$ to  $\mathbf{U}(t_{n+1})$ with $t_n$ the discrete time.  The spatial derivatives needed at each step of the RK algorithm are computed using a fourth order 11-point stencil finite-difference centered scheme \citep{bogey_family_2004} whose coefficients are optimized to minimize the dispersion error over a large range of wavenumber. 
For the points near the boundary, non-centered 11-points stencil schemes are used \citep{berland_high-order_2007}. 
Selective filters are also applied to remove grid-to-grid oscillations. 

\begin{figure}
    \centering
\includegraphics[]{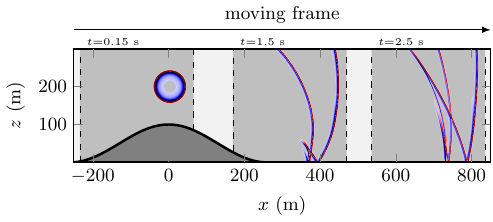}
\caption{Snapshots of the moving frame at three different instants in time as the wave propagates inside the domain.}
\label{fig:moving-frame}
\end{figure}

\paragraph{Moving frame.}
A moving frame method~\citep{dragna_towards_2014} is used to keep an affordable computational cost for long range propagation.
In this method, the distance between the wavefront and the right boundary of the domain is computed at each iteration and the acoustic variables are shifted to maintain this distance constant throughout the simulation (see Fig.~\ref{fig:moving-frame}). 
This approach allows for a large reduction of the computational domain size. Thus, in the simulations, a $300$~m$~\times~ 300$~m moving frame is used to calculate the sound propagation in a $3000$~m$~\times~ 300$~m domain, which reduces the computational cost by one order of magnitude. 
\paragraph{Boundary conditions}
A broadband impedance condition is used at the bottom boundary of the domain to model realistic ground absorption.
The impedance condition is developed in the time domain~\cite{rienstra_impedance_2006} such that it writes as a convolution.
By taking the surface impedance as a rational function of the frequency, it is possible to substitute this convolution by additional differential equations that can be solved along the RK scheme \cite{troian_broadband_2017}. This saves computational resources, while preserving high-order accuracy. In this work, the method was not implemented on the impedance but on the reflection coefficient for stability reasons.

At the top of the domain a perfectly matched layer (PML) is implemented to simulate unbounded atmosphere. 
Inside this layer, the derivative in $z$ in Eq.~(\ref{eq:conservative}) is modified such that
\begin{equation}
    \frac{\partial}{\partial z} \longrightarrow \frac{1}{\kappa+\sigma / i \omega} \frac{\partial}{\partial z},
\label{eq:pml}
\end{equation}
with
\begin{equation}
    \begin{aligned}
    &\sigma\left(z\right)=\sigma_0\left[\left(z-z_{\mathrm{PML}}\right) / L\right]^3,\\
    &\kappa\left(z\right)=1+\left(\kappa_0-1\right)\left[\left(z-z_{\mathrm{PML}}\right) / L\right]^3,
    \end{aligned}
\end{equation}
where $z_\mathrm{PML}$ is the coordinate at the bottom of the layer and $L$ is its thickness. 
As for the impedance condition, this transformation can be written in the time domain as a convolution that is integrated along the RK scheme with additional differential equations \citep{cosnefroy_simulation_nodate,komatitsch_unsplit_2007}. 

Finally, on the left side of the domain, the waves must also leave the computational domain without producing any reflections. 
To achieve this, the sound speed is gradually reduced in a thin layer to ensure that potentially reflected waves move slower than the domain, which is moving at the speed of sound. 
Hence, they can not re-enter the domain \citep{cosnefroy_simulation_nodate}.

\paragraph{Post-processing.}
During the computation, pressure time signals $p(t)$ are recorded at several receiver locations. 
Then, the Fourier transform $\hat{p}(\omega)$ of each signal is computed, and the relative sound pressure is retrieved by dividing $\hat{p}(\omega)$ by the solution for a Gaussian pulse in free field,
\begin{equation}\label{eq:deltaL_LEE}
    \Delta L(\mathbf{x},\omega) = 10\log_{10} \left(\frac{|\hat{p}(\mathbf{x},\omega)|^2}{|\hat{p}_{\mathrm{ff}}(\mathbf{x},\omega)|^2}\right),
\end{equation}
where $\hat{p}_{\mathrm{ff}}$ is given by: 
\begin{equation}\label{eq:pff_LEE}
    \hat{p}_{\mathrm{ff}}(\mathbf{x},\omega) = \frac{\pi k_0 B^2S_0}{4 c_0} \exp \left( \frac{-k_{\mathrm{eff}}^2B^2}{4} \right) H_0^{(1)}(k_0R),
\end{equation}
\replaced{
$k_{\mathrm{eff}}=k_0/(1+M)$ with $M=u_0/c_0$ the Mach number at the wind turbine's hub.
The term $k_{\mathrm{eff}}$ accounts for the effect of the flow on the source which induces a shift in frequency.}{
$H_0^{(1)}$ is the zeroth order Hankel function of the first kind, $k_0$ is the wave number, and $k_{\mathrm{eff}}=k_0/(1+M)$ with $M=u_0/c_0$ the Mach number at the wind turbine hub's height.
The term $k_{\mathrm{eff}}$ accounts for the effect of the flow on the source which induces a shift in frequency.
}
The derivation of this correction term can be found in the \replaced{}{Appendix}. 

Once $\Delta L$ is computed, it can be used in the SPL prediction methodology explained in Sec.~\ref{sec:general_layout}. 

\subsubsection{PE method}
\paragraph{PE formulations.}
Two 2D PE formulations are implemented.
The first one is a curvilinear formulation of the wide-angle PE with the effective sound speed approximation (\replaced{}{\GTPE}). 
It is derived from the methodology described by \citet{sack_parabolic_1995} \replaced{}{for the generalized terrain parabolic equation (GT-PE)}, but with the coordinate transformation shown in Eq.~(\ref{eq:curvi-transform}). 
The \replaced{}{\GTPE} allows us to consider topography with slope \replaced{up to}{that does not exceed around} $30^{\circ}$ \citep{salomons_computational_2001}.
\replaced{}{
The moving atmosphere is modeled  by defining an effective sound speed profile such that 
\begin{equation}
  c_{\rm eff}(x,z) = c_0(T_0(x,z)) + u_0(x,z),
\end{equation}
where $c_0$ is the sound speed that depends on the mean temperature $T_0$.
}
It is worth mentioning that the \replaced{GT-PE}{\GTPE} is equivalent to the classical effective sound speed WAPE formulation for flat ground. 
The accuracy of the \replaced{GT-PE}{\GTPE} strongly depends on quantities computed at the ground (ground profile and its derivatives, ground surface impedance...), where small numerical errors can appear and accumulate \citep{sack_parabolic_1995}. 
Errors can also arise from the effective sound speed approach with phase errors accumulating over the distance.  

\replaced{
The second PE is a wide angle vector PE formulation ($\mathrm{WAPE_{vec}}$), that incorporates a vector wind field (although only the horizontal component of the wind velocity $u_0$ is considered) and that is valid for arbitrary high Mach numbers. 
}{
The second PE is a wide-angle vector PE formulation ($\mathrm{WAPE_{vec}}$), that incorporates a vector wind field  and is valid for arbitrary high Mach numbers. 
}
\replaced{}{
It  is presented in Sec.~VI.~A of \citet{ostashev_wave_2020}.
Here the velocity field is considered with fewer assumptions.
However, it was not derived in curvilinear coordinates and hence can not consider topography. 
While WAPE$_{\rm vec}$ can consider both the vertical and horizontal components, only the horizontal component of the wind velocity $u_0$ is considered in our implementation.
}

The two methods employ second-order finite-difference schemes in the $z$-direction and a \replaced{Crank-Nicholson}{Crank-Nicolson} algorithm to advance the solution to $x+ \Delta x$ from the solution at $x$. Numerically, this involves inverting a tridiagonal matrix at each step of the resolution, which is done efficiently using the Thomas algorithm. The starting field is the second-order starter presented in \citet{salomons_computational_2001}\added{, which represents a monopole source}.

\paragraph{Boundary condition}
At the top of the domain a PML is implemented \citep{collino_perfectly_1997} by modifying the partial derivative along the vertical coordinate as for the LEE method [see Eq.~(\ref{eq:pml})]. 
At the bottom of the domain an impedance boundary condition is used. 
The implementation of the boundary conditions leads to modifications of the matrix coefficients, \replaced{}{while} conserving the tridiagonal shape.
The boundary conditions are similar for the two PE formulations but, because of the differences in the equations, the resulting modifications of the matrix coefficients are not identical.
\replaced{The derivation of the matrix formulation for the modified \replaced{GT-PE}{\GTPE} can be found in the supplementary material \footnote{See
Supplementary materials at [URL will be inserted by AIP]
for [Formulation of the modified generalized-terrain parabolic equation].}}{}.

\section{Cases studied}\label{sec:cases}
\begin{figure}[ht]
\centering
\includegraphics[]{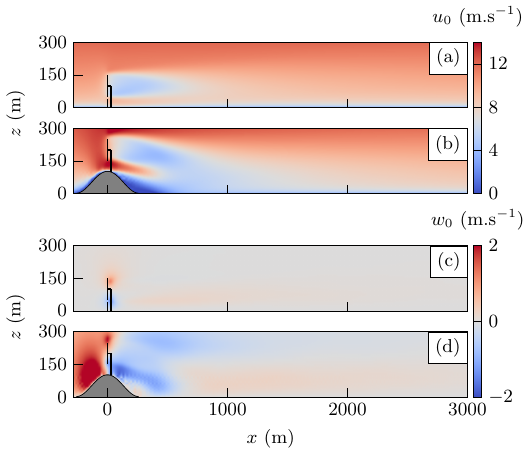}
\caption{Mean axial velocity field $u_0$ for the two cases of interest: (a) a wind turbine over flat ground, (b) on top of a hill. {\color{track} Mean vertical velocity field $w_0$ for (c) a wind turbine over flat ground, (d) on top of a hill.}}
\label{fig:flow}
\end{figure}
The study examines two scenarios: \replaced{a basic scenario}{one} with a wind turbine on a flat surface and \replaced{a scenario}{one} with a wind turbine positioned on a hilltop. The latter scenario is particularly relevant, as wind turbines are often installed on hills to maximize energy output. Therefore, it is crucial to accurately model sound propagation in such complex settings.
For both cases, LES results from \citet{liu_effects_2020}  for a truly neutral ABL are used. 
The mean flow fields from these simulations are normalized by the friction velocity \replaced{$u^*$}{$u_*$} and by the diameter of the wind turbine. 
In this work, the wind turbine hub height and diameter are both set to 100~m. The friction velocity is set to \replaced{$u^*$}{$u_*$}$=0.512~$m~s$^{-1}$ to get a wind velocity equal to $10~$m~s$^{-1}$ at the hub height.
For simplicity, the temperature and the sound speed are assumed constant, which is a valid assumption when considering a neutral atmosphere but would not be for a stable or unstable one.  
For the second case, the hill is defined such that:
\begin{equation}
    h(x) = h_{\max} \cos^2\left(\frac{\pi x}{2l}\right) \; {\rm for} \;  -l<x<l \; , 
\end{equation}
where $h(x)$ is the terrain elevation,  $h_{\max}=100$~m, and $l = 260~$m is the half-width of the hill.
The mean axial velocity component is plotted for both cases in Fig.~\ref{fig:flow}{\color{mblue}(a)} and \ref{fig:flow}{\color{mblue}(b)}.
A velocity deficit can be observed just after the turbine in both cases. 
In the first case, a ducting effect is expected due to the wind turbine wake~\citep{barlas_effects_2017}. 
The acoustic waves are trapped inside the wake and then redirected towards the ground when the unperturbed ABL is recovered.
This generates a focusing zone at the ground with high SPL. 
Ducting in the presence of the hill is expected to be stronger as the wake is more pronounced and more directed towards the ground. 
The vertical component of the wind speed, plotted Fig.~\ref{fig:flow}\textcolor{mblue}{c}, is small in the flat case (less than 4~\% of $u_0$), so that the vector PE and the LEE are expected to produce similar results.
In the case with topography, the hill induces a vertical component of the wind speed just before the wind turbine of a few meters per second, \replaced{}{shown in Fig.~\ref{fig:flow}\textcolor{mblue}{d}}, which is not taken into account by the \replaced{}{\GTPE}.
This difference of a few meters per second upstream of the hill could introduce some discrepancies \replaced{}{between the LEE and \GTPE} \replaced{}{methods}  in the final OASPL.
\replaced{}{In addition, errors could arise for both cases due to the one-way approximation and the angular validity of the two PE methods.}

The variable porosity model \cite{attenborough_outdoor_2011} is used with a flow resistivity of 50~kN~s~m$^{-4}$ and a porosity change rate of 100~m$^{-1}$ to model a grassy ground.
The PE simulations are performed for a set of frequencies used to compute third octave band spectrum between 50~Hz and 1~kHz. 
Values of $\Delta L$ for the same frequencies are extracted from the broadband results of the LEE simulation for comparison.
These frequencies are gathered in  Table~\ref{tab:frequ}.
\begin{table}[ht!]
  \caption{{\color{track} Frequencies $f$ used to compute the third octave band spectrum with the PE methods. $f_c$ is the central frequency of each band.}}\label{tab:frequ}
  \begin{ruledtabular}
    \scriptsize
  \begin{tabular}{ccccccccccccccc}
  $f_c$ (Hz) &50&63&80&100&125&160&200&250&315&400&500&630&800&1000\\
  \hline
  $f$ (Hz)&50&63&80&100&125&160&192&241&297&373&467&588&741&926\\
  & & & & & & &   208  & 260 & 315 & 391 & 489 & 616 & 770 & 962\\
  & & & & & & &     &     & 334 & 409 & 512 & 645 & 800 & 1000\\
  & & & & & & &     &     &    & 429 & 536 & 675 & 831 & 1039\\
  & & & & & & &     &     &    &     &     &     & 864 & 1080\\
  \end{tabular}
  \end{ruledtabular}
  \end{table}
For all simulations, the numerical parameters are set to obtain accurate results up to 1~kHz and 3~km downwind of the turbine. 
For the LEE, this implies setting a grid size of 0.05~m and a CFL of 0.5, which corresponds to 127000 time iterations to complete the simulation with a computational moving domain of $36 \times 10^{6}$ points. 
The computational time for one simulation is approximately 1200 central processing units (CPU) hours, which is equivalent to 3 days on a 16-core machine. 
For $\mathrm{WAPE}_{\mathrm{vec}}$ and \replaced{GT-PE}{\GTPE}, the grid size depends on the computed frequency.
For the flat case convergence was reached for $\Delta x = \Delta z = \lambda/10$, which requires 1~CPU hour to compute the results at all frequencies.
The hill case requires a higher resolution to capture sharp gradients and convergence is reached for  $\Delta x = \Delta z = \lambda/50$, which leads to an increased computational time of 30~CPU~hours.

The source model used for the wind turbine is the same as described in \citet{tian_wind_2016}. The only difference is that the wind turbine is scaled up to be of 100~m in diameter instead of 93~m. 
The blade is decomposed into eight segments, and 36 angular positions are considered for one rotation of the blades. 
Finally, it was found by \citet{cotte_extended_2019} that seven fictive source heights are sufficient  to obtain a convergence on SPL and AM for the flat case at ground level using the approach described in Sec.~\ref{sec:general_layout}. 
However, for the case with a hill, 30 source heights are required for convergence.
This significant increase in the number of sources needed to achieve convergence can be explained by the strong dependence between the focusing pattern and the source height in the presence of the hill. 
Thus, changing the source height by a few meters can affect the position of the focusing zone at the ground of more than 100~m. 
Therefore, it is necessary to reduce the step between the simulated source heights. 

\section{Comparison}\label{sec:results}
\subsection{Over flat ground}
\subsubsection{Relative SPL}
\begin{figure*}[]
\includegraphics[]{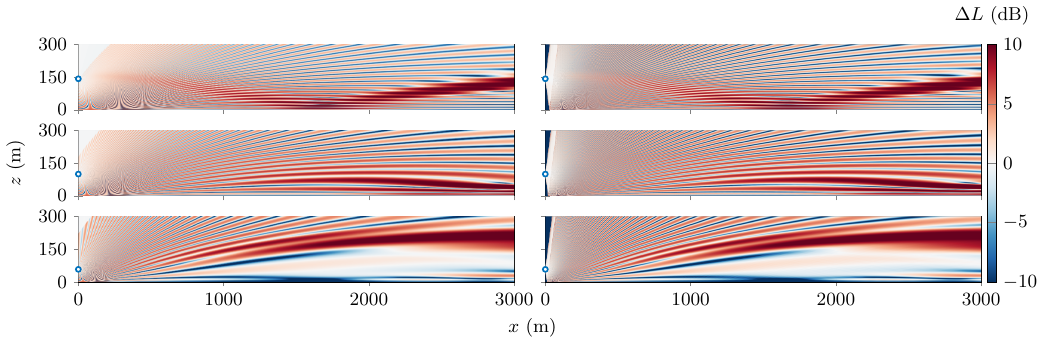}
\caption[]{$\Delta L$ computed with LEE (left) and $\mathrm{WAPE}_{\mathrm{vec}}$ (right)  at 100~Hz for several source heights (circles): 142 m (top), 100 m (middle), and 58 m (bottom).}
\label{fig:deltaLcartoAb}
\end{figure*}

\begin{figure*}[ht]
  \includegraphics[]{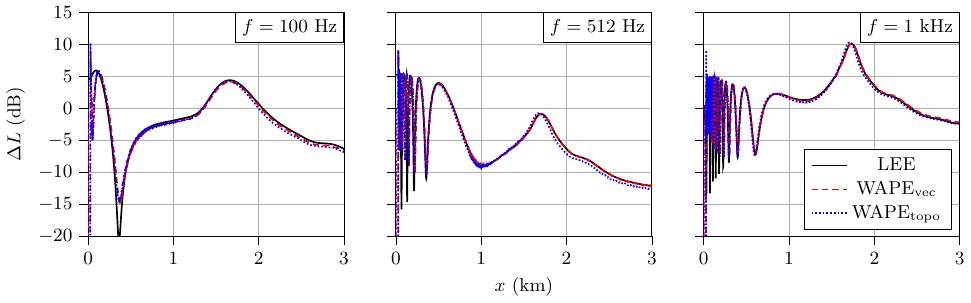}
  \caption[]{$\Delta L$ computed for a source at 142~m and a receiver at 2~m for three different frequencies.}
  \label{fig:deltaLlineAb}
\end{figure*}
The value of $\Delta L$ represents the effect of the ground absorption and \replaced{of}{} the mean flow on the propagation with respect to the solution in the free field.
The $\Delta L$ fields are depicted in Fig.~\ref{fig:deltaLcartoAb} for LEE and $\mathrm{WAPE}_{\mathrm{vec}}$ at 100~Hz for three source heights corresponding to the highest (142~m) and lowest (58~m) fictive sources and to the fictive source at hub height (100~m). 
The results for the \replaced{GT-PE}{\GTPE} are not plotted as they are almost identical with those of the $\mathrm{WAPE}_{\mathrm{vec}}$. 
In this case, the two main effects are the occurrence of constructive and destructive interference patterns due to ground reflection and the ducting of acoustic waves by the wake of the wind turbine. 
In particular, note the large increase in  $\Delta L$  at $x=1.7$~km at the ground for the source at 142~m height. 
It is clear that the same interference patterns and the same ducting are obtained in the $\mathrm{WAPE}_{\mathrm{vec}}$ and LEE results. 
The influence of the \replaced{wake}{wind speed gradient and of the wake length and intensity} on the propagation for different source heights is well captured with both methods \replaced{and}{as} the  $\Delta L$ levels are similar. 
Still, a small difference between the two methods can be observed at the very beginning of the domain ($x<100~$m).
For the LEE method, the moving frame does not allow the reflected wave to reach the top of the domain close to the source, leading to a zone without any interferences and a value of $\Delta L$ close to 0. 
In the case of PE, the classic cone due to the angle of validity of the method can be seen at the very beginning of the domain.
Hence, the two methods do not produce the same results in the near field.

The relative sound pressure level is plotted for a line of receivers 2~m above the ground in Fig.~\ref{fig:deltaLlineAb} for three different frequencies. 
At all frequencies, a strong peak  is visible at 1.7~km which corresponds to the distance where the focused wave hits the ground. 
The values of $\Delta L$ obtained with \replaced{GT-PE}{\GTPE} and $\mathrm{WAPE}_{\mathrm{vec}}$ methods are compared with the LEE results.
The curves are almost identical for the three methods.
\replaced{
}{The difference between the $\Delta L$ obtained with the three methods does not exceed 0.5~dB for $x>1~$km.
}At $512~$Hz and $1000~$Hz a small shift is found between the  position of the peaks in the \replaced{GT-PE}{\GTPE} and $\mathrm{WAPE}_{\mathrm{vec}}$ results.
The latter seems, as expected, to better account for the wind field and the results are closer to those of the LEE. 
For other source or receiver heights (not plotted here), it was observed that the difference between \replaced{GT-PE}{\GTPE} and LEE results tends to increase with distance. 
It was also seen that as the interference pattern gets more complex, the difference is more visible.

\begin{figure}[]
  \centering
  \includegraphics[]{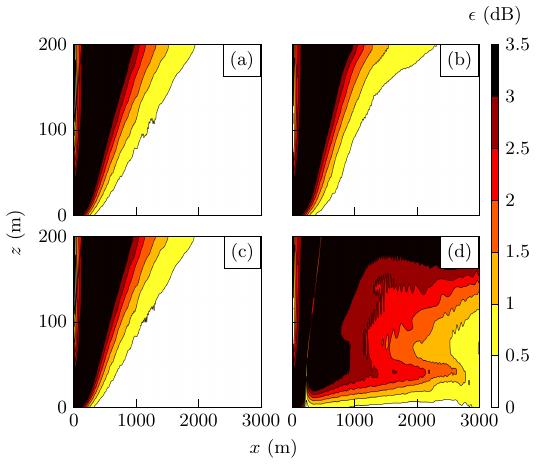}
  \caption[]{$\Delta L$ difference computed between the LEE and (a,b) WAPE$_{\rm vec}$ and (c,d) {\color{track}\GTPE} for (a,c) an atmosphere at rest and (b,d) for an ABL profile with 10~m~s$^{-1}$ at hub height. The results are averaged over $N_f = 35$ frequencies, see Table~\ref{tab:frequ}, {\color{track}and are computed for flat terrain.}}
  \label{fig:errorAb}
  \end{figure}
    
The average difference of $\Delta L$ is plotted in Fig.~\ref{fig:errorAb} to quantify the error made with PE formulations. This difference is defined with:
\begin{equation}
 \epsilon = \frac{1}{N_f}\sum_{i=1}^{N_f} |\Delta L_{\mathrm{LEE}}(f_i) - \Delta L_{\mathrm{PE}}(f_i)| \;,
\end{equation}
with $N_f$ the total number of frequencies computed. 
In order to assess the effect of the mean flow on the results, an additional simulation for an atmosphere at rest  is performed for all three methods. 
The difference for the atmosphere at rest is shown in Fig.~\ref{fig:errorAb}\textcolor{mblue}{a} and \ref{fig:errorAb}\textcolor{mblue}{c}.
Note that the error is almost identical for the $\mathrm{WAPE}_{\mathrm{vec}}$ and the \replaced{GT-PE}{\GTPE}.\@
The error is very large close to source due to the angle of validity of the PE \replaced{}{and can even go above 3~dB}.\@
Further away, the propagation angle lies within the angle of validity of the \replaced{GT-PE}{\GTPE}, and the error reduces to less than 0.5~dB. 
The introduction of a mean flow does not significantly modify the error for the $\mathrm{WAPE}_{\mathrm{vec}}$ (Fig.~\ref{fig:errorAb}\textcolor{mblue}{b}). 
The effect of the flow is very well accounted for in this method, as expected. 
On the contrary, for \replaced{GT-PE}{\GTPE} the addition of a mean flow (Fig.~\ref{fig:errorAb}\textcolor{mblue}{d}) greatly increases the error. 
The difference  is not only significant in the near field but in the entire domain with differences going \replaced{}{above} 3~dB. 
It can be observed that close to the ground the error is still relatively small, which corresponds to the case presented in Fig.~\ref{fig:deltaLlineAb}.
It is worth noting that, because we are considering a flat ground, the error in the \replaced{GT-PE}{\GTPE} method is only due to the effective sound speed approach. 
\subsubsection{SPL}
\begin{figure}
\centering
\includegraphics[]{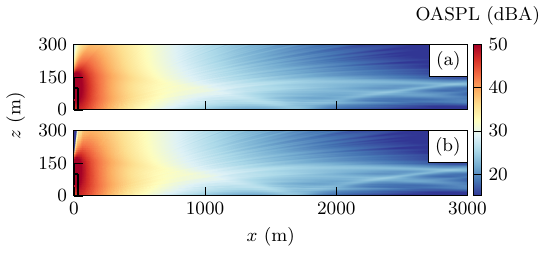}
\caption{{\color{track}OASPL at $\beta=0^\circ$ for the flat case computed using (a) LEE  and (b) \GTPE}.}\label{fig:oasplCartoAb}
\end{figure}
In this section, the SPL obtained by combining $\Delta L$ with the source model described in Sec.~\ref{sec:general_layout} are compared for the LEE and the two PE methods. 
The OASPL field for one angular position of the blades obtained from LEE and \replaced{GT-PE}{\GTPE} methods is presented in Fig.~\ref{fig:oasplCartoAb}. 
It corresponds to the contribution of all blade segments for one angular position (here $\beta=0^\circ$, i.e. for one of the three blades pointing upwards) summed over the frequency bands between 50~Hz and 1~kHz.
The results from $\mathrm{WAPE}_{\mathrm{vec}}$ are omitted as they are again almost identical with those of \replaced{GT-PE}{\GTPE}.
Several zones of large OASPL produced by the different sources distributed along the blades can be observed in both cases. The OASPL obtained from the \replaced{GT-PE}{\GTPE} and the LEE methods for several positions of the blades is available in video Mm.~\ref{mmtest1}.
As the position of the blades changes, different $\Delta L$ calculations are activated, leading to distinct focusing zones on the ground. 
This mechanism is responsible for the amplitude modulation in the far field \citep{barlas_variability_2018}.
Hence, it is not expected that strong discrepancies appear as the error observed in Fig.~\ref{fig:errorAb} would tend to average out with an extended broadband source. 
\multimedia{../figures_offline/Ab/compar/film/snap.mp4}{Evolution of the OASPL with the angle $\beta$,  in the flat case for the \replaced{GT-PE}{\GTPE} and the LEE methods. 
}\label{mmtest1}
\begin{figure}[]
  \includegraphics[]{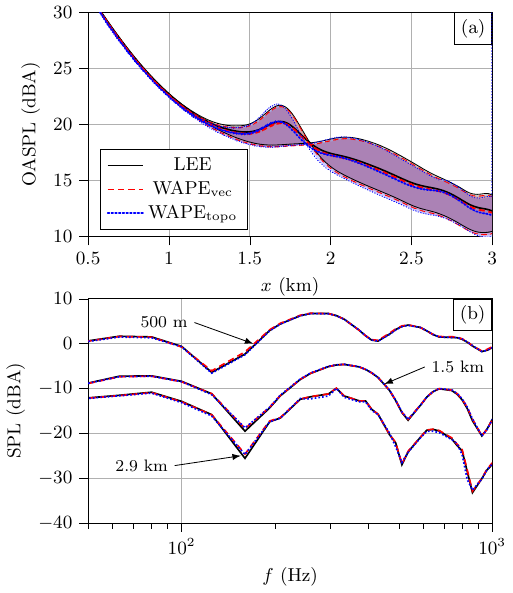}
  \caption{ (a) Averaged OASPL over one rotation computed with the three propagation methods, minimum and maximum delimited with color patches. (b) {\color{track}
    Averaged spectrum obtained over a full rotation of the turbine blades for receivers at a height of 2~m, positioned at distances of $x=500~$m, $x=1.5~$km, and $x=2.9~$km downstream of the turbine (listed from top to bottom).
  }}\label{fig:splLineAb}
  \end{figure}

The mean OASPL over one rotation \replaced{}{for receivers located 2~m above the ground} is shown in Fig.~\ref{fig:splLineAb}\textcolor{mblue}{a}. 
The minimum and maximum values reached during the rotation are also delimited by color patches.
The mean OASPL obtained with the three methods is very similar.
The peak observed 1.7~km from the source is well captured by both PE methods.
The small phase shift in $\Delta L$ observed for the \replaced{GT-PE}{\GTPE} is still present in the OASPL prediction but is less pronounced. 
The two amplitude modulation zones from 1.25~km to 1.9~km and from 1.9~km to 3~km are similar with all methods. It corresponds to the areas where the SPL varies due to the motion of the blades.  

The narrow band spectrum at three downstream locations is shown in Fig.~\ref{fig:splLineAb}\textcolor{mblue}{b} for the LEE and the two PE formulations.
\replaced{}{Spectra are nearly identical over the considered frequency range.}
The dips induced by ground absorption are equally captured.
The error does not increase with frequency or distance, which shows the good convergence of the three methods for this frequency range and domain length.

\subsection{On top of a hill}
\subsubsection{Relative SPL}
\begin{figure*}[]
\includegraphics[]{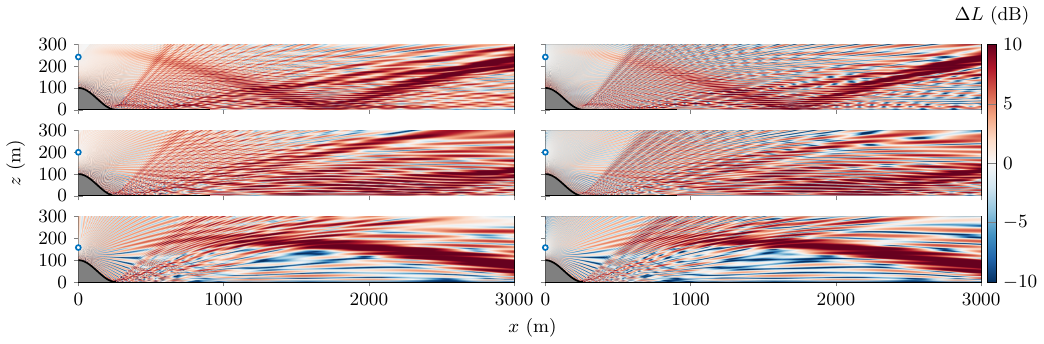}
\caption[]{$\Delta L$ computed with LEE (left) and {\color{track}\GTPE} (right)  at 100~Hz for several source heights (circles): 242 m (top), 200 m (middle), and 158 m (bottom).}
\label{fig:deltaLcartoCb}
  \end{figure*}
\begin{figure*}[]
\includegraphics[]{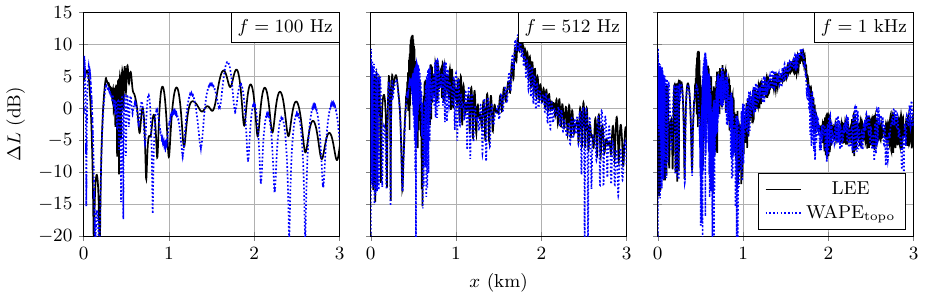}
\caption[]{$\Delta L$ computed for a source at 242~m and a receiver at 2~m for three different frequencies. }
\label{fig:deltaLlineCb}
\end{figure*}
 
The presence of the hill influences sound propagation  by its geometry and  by inducing a more complex mean flow, which changes the refraction of acoustic waves in the ABL. 
This refraction pattern is shown in Fig.~\ref{fig:deltaLcartoCb} for the LEE and the \replaced{GT-PE}{\GTPE} methods and for three source heights.  
As for the flat case, different focusing zones can be identified depending on the source height.
The presence of a caustic at the bottom of the hill can also be observed.
The interference dips look more pronounced for the \replaced{GT-PE}{\GTPE} and the levels are slightly higher for the LEE.
Furthermore, important differences can be seen in the results, especially at 100~Hz, for a line of receivers at 2~m height and for a source at 242~m (Fig.~\ref{fig:deltaLlineCb}).
These large discrepancies are due to a combined effect of the hill and of the mean flow. 
In fact, the steep slope of the hill is the main cause of the error in this case as discrepancies were already visible without any mean flow (not shown here). The maximum slope of the hill (reached at $x =130$~m) is equal to $31^\circ$, which is at the limit of the \replaced{GT-PE}{\GTPE} validity range \citep{salomons_computational_2001}.
It is worth noting that the regularity of the ground profile or the precision with which its derivatives are calculated are not an issue here as they are analytical. 
The mean error in the whole domain for all frequencies is between 3~dB and 4~dB for all heights, which is 2~dB higher than for the flat case. 
Still it can be observed that the effect of the hill and the mean flow is the same for the LEE and \replaced{GT-PE}{\GTPE}. Thus, the shape of the peak at 1.8~km is similar for both methods at 512~Hz and 1000~Hz. 
Even if the \replaced{GT-PE}{\GTPE} introduces errors in this case, the effects of the ABL and the hill remain well captured both in terms of amplitude and position of the focusing. 
 
\subsubsection{SPL}
\begin{figure}[h]
  \centering
  \includegraphics[]{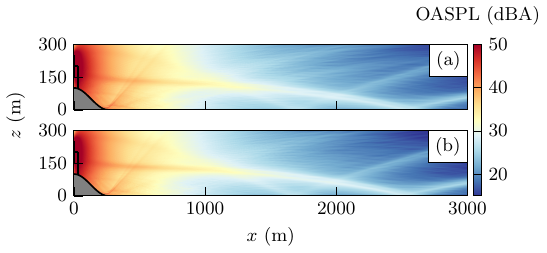}
  \caption[]{OASPL at $\beta=0^\circ$ for the hill case computed using (a) LEE  and (b) {\color{track}\GTPE}.}\label{fig:oasplCartoCb}
  \end{figure}
As for the flat \replaced{}{terrain} case, the OASPL map depicted in Fig.~\ref{fig:oasplCartoCb} corresponds to the superposition of SPL from the different sources distributed along the blades. 
The caustic observed at the bottom of the hill in Fig.~\ref{fig:deltaLcartoCb} is still present, as well as several sound focusing zones.  
A notable difference from the flat \replaced{}{terrain} case is that the focusing is significantly stronger. 
Additionally, as discussed in Sec. \ref{sec:cases}, the presence of the hill results in greater amplitude modulation due to blade movement.
This can be observed in Mm. \ref{mmtest2}, and it is clear that both methods capture this phenomenon.
\multimedia{../figures_offline/Ab/compar/film/snap.mp4}{Evolution in the hill case of the OASPL with the angle $\beta$ for the \replaced{GT-PE}{\GTPE} and the LEE method. 
}\label{mmtest2}
\begin{figure}[h]
  \includegraphics[]{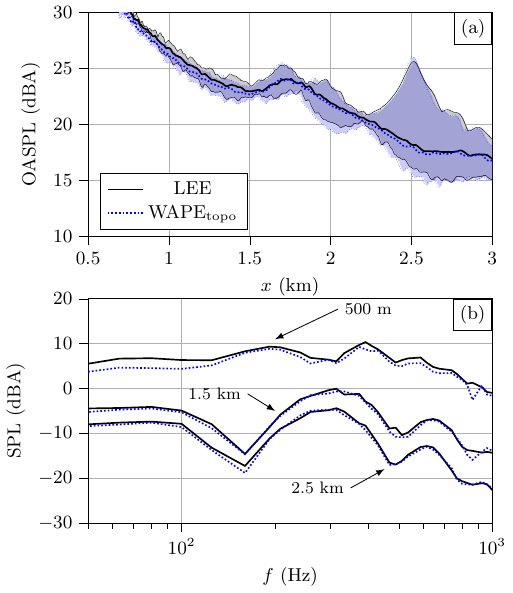}
  \caption{(a) Averaged OASPL over one rotation computed with the LEE and {\color{track}\GTPE} methods, minimum and maximum delimited with color patches. (b) {\color{track}
  Averaged spectrum obtained over a full rotation of the turbine blades for receivers at a height of 2~m, positioned at distances of $x=500~$m, $x=1.5~$km, and $x=2.5~$km downstream of the turbine (listed from top to bottom).
  }}\label{fig:splLineCb}
  \end{figure}  
To further investigate the discrepancies between the LEE and \replaced{GT-PE}{\GTPE} methods, the mean OASPL over one rotation is plotted in Fig.~\ref{fig:splLineCb}\textcolor{mblue}{a} along with the minimum and maximum OASPL for a line of receivers at 2~m height. 
The burst in mean OASPL at 1.8~km is similar for the two methods  as well as the maximum of amplitude modulation with a strong peak at 2.5~km. 
Hence, it is clear that the same propagation effects are captured with both methods. 
However, the OASPL computed with LEE are slightly higher than those computed with the \replaced{GT-PE}{\GTPE} (less than 1~dB).
Finally SPL spectra at three receiver positions are presented in Fig.~\ref{fig:splLineCb}\textcolor{mblue}{b}. 
\replaced{They show that \replaced{GT-PE}{\GTPE} slightly underpredicts SPL at low frequency for the receiver close to the source but provides an excellent agreement with LEE over the entire frequency range for the  other two receivers. }{
  The figure demonstrates that \GTPE~and LEE exhibit excellent agreement over the entire frequency range at $x=1.5$~km and $x=2.5$~km downstream of the turbine. 
  However, for the receiver situated close to the turbine (at 500 meters downstream), \GTPE~slightly underpredicts the SPL compared to LEE.
}

\section{Conclusion}\label{sec:conclusion}

The use of a new method based on the linearized Euler equations (LEE) for predicting wind turbine noise propagation was investigated.
The method includes a flow model based on LES and an extended source model based on Amiet's theory. Comparison with the state-of-the-art PE methods ($\mathrm{WAPE}_{\mathrm{vec}}$ and \replaced{GT-PE}{\GTPE}) was performed for two cases: a baseline case with a wind turbine over flat ground and a more complex case where the wind turbine  is positioned at the top of a hill. 
For the baseline case, the sound pressure level relative to the free field and the OASPL are almost identical when using LEE and the $\mathrm{WAPE}_{\mathrm{vec}}$. 
It was noticed that, as expected, the \replaced{GT-PE}{\GTPE} introduces phase errors due to the effective sound speed approach.
However, this has a minor impact on the prediction of the sound pressure levels at the ground and of OASPL.
\replaced{}{It is worth noting that the WAPE$_{\rm vec}$ is still preferable as it is derived with less assumptions and that it could be improved further by considering the vertical component of the wind speed.}
In the case with topography, the \replaced{GT-PE}{\GTPE} generates a noticeable error in the relative sound pressure level computed for a given frequency (of the order of 3~dB). This is due to the steep slope of the hill. Nevertheless, the OASPL and AM obtained with the \replaced{GT-PE}{\GTPE} remain very close to those obtained with the LEE method. 
The main effects of the flow on the propagation are still well simulated by the \replaced{GT-PE}{\GTPE}, and that the shifts in the interference pattern observed at each frequency tend to average out for the prediction of overall levels. 

Hence, we find that for situations with topography  PE methods in general and \replaced{GT-PE}{\GTPE} \replaced{}{in particular provide a suitable first approach to determine sound propagation from wind turbines.}.  
It is still worth noting the advantages of using LEE.
First, a time-domain solution is obtained  which allows one to compute a broadband SPL spectrum. The flow is taken into account with fewer assumptions and higher wind speed can be considered without introducing errors due to the effective sound speed approach. The main drawback of this method is its computation cost, especially when numerous source heights must be considered.


\begin{acknowledgments}
  The authors thank Luoqin Liu for providing access to the LES data of \citet{liu_effects_2020}.
  This work was performed within the framework of the LABEX CeLyA (ANR-10-LABX-0060) of Universit\'e de Lyon, within the program ``Investissements d’Avenir" (ANR-16-IDEX-0005) operated by the French National Research Agency (ANR).
  The authors were granted access to the HPC resources of PMCS2I (P\^ole de Mod\'elisation et de Calcul en Sciences de l'Ing\'enieur et de l'Information) of Ecole Centrale de Lyon, PSMN (P\^ole Scientifique de Mod\'elisation Num\'erique) of ENS de Lyon and P2CHPD (P\^ole de Calcul Hautes Performances D\'edi\'es) of Universit\'e Lyon I, members of FLMSN (F\'ed\'eration Lyonnaise de Mod\'elisation et Sciences Num\'eriques), partner of EQUIPEX EQUIP@MESO.
  This work was supported by the Franco-Dutch Hubert Curien partnership (Van Gogh Programme No. 49310UM).
  For the purpose of Open Access, a CC-BY public copyright license has been applied by the authors to the present document and will be applied to all subsequent versions up to the Author Accepted Manuscript arising from this submission.
  \replaced{}{
  The authors declare no conflicts of interest.
  Data from numerical simulations are available from the authors on reasonable request.
  }
\end{acknowledgments}
  

\appendix*
\section{Calculation of the noise level relative to the free field for the LEE}

The appendix details the derivation of the formula in Eq.~(\ref{eq:deltaL_LEE}) for calculating the noise levels relative to the free-field from the LEE solution. 

\subsection{Analytical derivation}

\begin{figure}[h]
	\centering
	\includegraphics[width=0.4\textwidth]{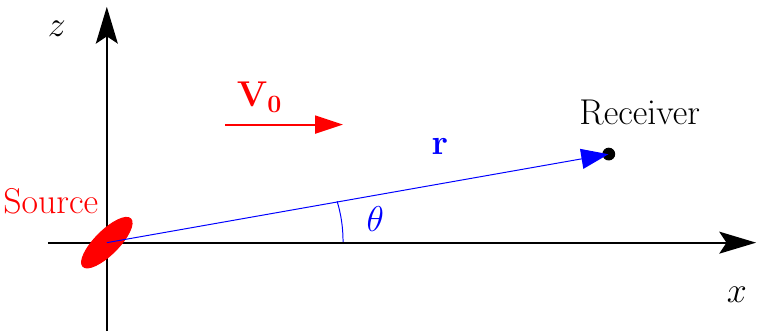}  
	\caption{Sound propagation from a spatially distributed source in a 2-D homogeneous uniformly moving medium.}
	\label{fig:sketch}
\end{figure}

We first derive the analytical solution in the frequency domain and in far-field for an impulsive source in a 2-D homogeneous uniformly moving medium and in free-field, as sketched in Fig.~\ref{fig:sketch}. The derivation closely follows that presented in~\citet{ostashev_equations_2005} for the case of a monochromatic point source.

\subsubsection*{Solution as a convolution}

A Cartesian system of coordinates $\mathbf{x} =(x,z)$ is used. 
A moving homogeneous atmosphere with a constant mean flow $\mathbf{V_0} = (u_0,0)$ is considered. The initial conditions are  $p(\mathbf{x},t=0)=S(\mathbf{x})$ and $\mathbf{v}(\mathbf{x},t=0)=\mathbf{0}$ and the source spatial distribution $S(\mathbf{x})$ is centered at the origin. Incorporating the initial conditions as source terms, the LEE in Eq.~(\ref{eq:LEE}) are written as:
\begin{align} \label{LEE_p.eq}
&\dfrac{\partial p}{\partial t} + u_0\dfrac{\partial p}{\partial x} + \rho_0c_0^2 \nabla \cdot \mathbf{v} =  S(\mathbf{x})\delta (t),\\ \label{LEE_v.eq}
&\dfrac{\partial \mathbf{v}}{\partial t} + u_0\dfrac{\partial \mathbf{v}}{\partial x} + \dfrac{1}{\rho_0}\nabla p = 0,
\end{align}
with $\delta$ the Dirac delta function. Combining Eqs.~(\ref{LEE_p.eq}) and (\ref{LEE_v.eq}) leads to:
\begin{equation} \label{onde.eq}
\left(\dfrac{\partial }{\partial t} + u_0\dfrac{\partial }{\partial x}\right)^2p - c_0^2 \Delta p = \left(\dfrac{\partial }{\partial t} + u_0\dfrac{\partial }{\partial x}\right)S(\mathbf{x})\delta (t).
\end{equation}
To translate the problem into the frequency domain, we employ the Fourier transform:
\begin{equation}
\hat{p}(\mathbf{x},\omega) = \int_{-\infty}^{\infty}p(\mathbf{x},t)\,{\rm e}^{{\rm i}\omega t}\,{\rm d}t.
\end{equation}
Taking the Fourier transform of Eq.~(\ref{onde.eq}) and dividing by $c_0^2$ gives:
\begin{equation}\label{p_hat.eq}
\Delta \hat{p} - \left(-{\rm i}k_0 + M\dfrac{\partial }{\partial x}\right)^2\hat{p} = -\dfrac{1}{c_0}\left(-{\rm i}k_0 + M\dfrac{\partial }{\partial x}\right)S(\mathbf{x})
\end{equation}
with $k_0 = \omega/c_0$ and $M = u_0/c_0$. 
Substituting in Eq.~(\ref{p_hat.eq})  the function $\hat{\phi}$ defined by:
\begin{equation}\label{phi_def.eq}
\hat{p} = -\dfrac{1}{c_0}\left(-{\rm i}k_0 + M\dfrac{\partial }{\partial x}\right)\hat{\phi}
\end{equation}
leads to:
\begin{equation}\label{phi_hat.eq}
\Delta \hat{\phi} - \left(-{\rm i}k_0 + M\dfrac{\partial }{\partial x}\right)^2\hat{\phi} = S(\mathbf{x}).
\end{equation}
For simplifying the previous equation, we use the transformation of coordinates $\mathbf{X} = (X, Z)$, with  $x = X/\gamma$, $z=Z$ and $\gamma = 1/\sqrt{1- M^2}$. Finally, introducing the function $\hat{\psi}$ defined by $\hat{\phi} =\exp(-{\rm i}K_0MX)\hat{\psi}$ with $k_0 = K_0/\gamma$ in Eq.~(\ref{phi_hat.eq}) yields:
\begin{equation}\label{psi_hat.eq}
\left(\dfrac{\partial^2 }{\partial X^2}+ \dfrac{\partial^2}{\partial Z^2} + K_0^2\right)\hat{\psi} = Q(\mathbf{X}).
\end{equation}
with $Q(\mathbf{X}) = \exp({\rm i}K_0MX)S(\mathbf{X})$. Eq.~(\ref{psi_hat.eq}) corresponds to the 2D inhomogeneous Helmholtz equation. Its solution is written as a convolution of the source term $Q(\mathbf{X})$ and the Green's function, which gives:
\begin{equation} \label{convolution.eq}
\hat{\psi}(\mathbf{X}) = -\dfrac{{\rm i}}{4}\int H_0^{(1)}(K_0|\mathbf{X}-\mathbf{X'}|)Q(\mathbf{X'})\,  {\rm d} \mathbf{X'},
\end{equation}
 with $H_0^{(1)}$ the zeroth order Hankel function of the first kind.

\subsubsection*{Far-field approximation}

In order to evaluate the convolution in Eq.~(\ref{convolution.eq}), a  far-field approximation, known as the Fraunhofer approximation, is performed. It is assumed that the source-receiver distance is large compared to the characteristic size of the source $B$, i.e. $|\mathbf{X}|\gg B$.  As a consequence, the source-receiver distance is approximated at the zeroth order in the amplitude of the integrand, i.e. $|\mathbf{X}-\mathbf{X'}|=|\mathbf{X}|$, and at the first order in its phase, i.e. $|\mathbf{X}-\mathbf{X'}| = |\mathbf{X}| - (\mathbf{X}\cdot \mathbf{X'})/|\mathbf{X}|$. Under the Fraunhofer approximation, one has:
\begin{multline}
H_0^{(1)}(K_0|\mathbf{X}-\mathbf{X'}|) = H_0^{(1)}(K_0|\mathbf{X}|)\exp(-{\rm i}K_0|\mathbf{X}|)\\ \exp\left[{\rm i}K_0\left(|\mathbf{X}| - \dfrac{\mathbf{X}\cdot \mathbf{X'}}{|\mathbf{X}|}\right)\right],
\end{multline}
which allows us to express the function $\hat{\psi}$ in Eq.~(\ref{convolution.eq}) as:
\begin{multline}
\hat{\psi}(\mathbf{X})= \dfrac{-{\rm i}}{4}H_0^{(1)}(K_0|\mathbf{X}|)\int S(X',Z')\\
\exp[-{\rm i}K_0(\cos\Theta -M)X']\exp(-{\rm i}K_0\sin\Theta Z')\,  {\rm d} X' {\rm d} Z'
\end{multline}
Returning in the physical space, we get:
\begin{multline} \label{psi_end.eq}
\hat{\psi}(\mathbf{x})= \dfrac{-{\rm i}\gamma}{4}H_0^{(1)}(k_0\gamma\sqrt{\gamma^2x^2 +z^2})\int S( x',z')\\
\exp[-{\rm i}k_0\gamma^2(\cos\Theta -M) x']\exp(-{\rm i}k_0\gamma\sin\Theta z')\,  {\rm d} x' {\rm d} z',
\end{multline}
where $\cos\Theta = X/|\mathbf{X}|$ and $\sin\Theta = Z/|\mathbf{X}|$ are related to $\cos\theta = x/|\mathbf{x}|$ and $\sin\theta = z/|\mathbf{x}|$ by:
\begin{align}
\cos \Theta =\dfrac{\cos\theta}{\sqrt{1 -M^2\sin^2\theta}}, \quad
\sin \Theta =\dfrac{\sin\theta}{\gamma\sqrt{1 -M^2\sin^2\theta}}.
\end{align}
Using Eq.~(\ref{psi_end.eq}), we can write an expression for the function $\hat{\phi}$ in compact form as:
\begin{multline} \label{phi_end.eq}
\hat{\phi}(\mathbf{x}) = \dfrac{-{\rm i}\gamma}{4}{\rm e}^{{\rm i}k_0\gamma^2Mx}H_0^{(1)}(k_0\gamma\sqrt{\gamma^2x^2 +z^2})\\
\tilde{S}(k_0\gamma^2(\cos\Theta -M),k_0\gamma\sin\Theta)
\end{multline}
with $\tilde{S}$ the spatial Fourier transform of $S(\mathbf{x})$:
\begin{equation}
\tilde{S}(k_x,k_z) =\int S(x,z)\,{\rm e}^{-{\rm i}k_x x-{\rm i}k_z z}\,  {\rm d} x\, {\rm d} z.
\end{equation}
Finally, the pressure in the frequency domain is obtained from Eqs.~(\ref{phi_def.eq}) and (\ref{phi_end.eq}):
\begin{multline} \label{p_end.eq}
\hat{p}(\mathbf{x}) = A(\omega,M,\theta)\dfrac{-{\rm i}\gamma^3}{4}{\rm e}^{{\rm i}k_0\gamma^2Mx}\left(H_0^{(1)}(k_0\gamma\sqrt{\gamma^2x^2 +z^2})\right.\\ \left.-
\dfrac{{\rm i}\gamma x M}{\sqrt{\gamma^2x^2 +z^2}}H_1^{(1)}(k_0\gamma\sqrt{\gamma^2x^2 +z^2})\right)
\end{multline}
where $A(\omega,M,\theta)$ represents the equivalent amplitude of the impulsive source: 
\begin{equation} \label{amplitude.eq}
A(\omega,M,\theta)=\dfrac{{\rm i} k_0}{c_0} \tilde{S}(k_0\gamma^2(\cos\Theta -M),k_0\gamma\sin\Theta).
\end{equation}

\subsection{Source correction}

At long range, we can assume that the acoustic waves have been launched at small angles, corresponding to $\theta\approx 0$. Denoting $A_M(\omega) \equiv A(\omega,M,\theta\approx0)$, the source amplitude becomes:
\begin{equation} \label{AM_def.eq}
A_M(\omega) = \dfrac{{\rm i} k_0}{c_0} \tilde{S}\left(\dfrac{k_0}{1+M},0\right).
\end{equation}

\subsubsection*{Case of a Gaussian spatial distribution}

In this work, the impulsive source has a Gaussian spatial distribution 
\begin{equation}
S(x,z) = S_0\exp\left(-\dfrac{x^2+z^2}{B^2} \right),
\end{equation}
whose Fourier transform is:
\begin{equation}
\tilde{S}(k_x,k_z) = \pi B^2 S_0 \exp\left(-\dfrac{(k_x^2+k_z^2)B^2}{4} \right).
\end{equation}
The source amplitude in Eq.~(\ref{AM_def.eq}) is then given by:
\begin{equation}\label{AM_Gauss.eq}
A_M(\omega) = \dfrac{{\rm i} k_0}{c_0}\pi B^2 S_0\exp\left(-\dfrac{k_0^2B^2}{4(1+M)^2} \right).
\end{equation}
Note that for $M=0$, the amplitude for the Gaussian impulse source in Eq.~(\ref{AM_Gauss.eq}) leads to the same expression as given in Eq.~(23) in~\citet{dragnaTimeDomainSimulationsOutdoor2011b} for the case of a homogeneous atmosphere at rest. 
\replaced{}{
  From this expression we can also assess the frequency content of the Gaussian pulse. 
  The maximum is reached for $f = c_0/(\sqrt{2} \pi B)$ and a cut-off frequency $f_{10}$ can be define such that the amplitude is equal to $10\%$ of this maximum: 
  \begin{equation}\label{eq:f_content}
    f_{10} \approx~0.6c_0/B.
  \end{equation}
}
\begin{figure}[h]
	\centering
	 \includegraphics[width=0.5\textwidth]{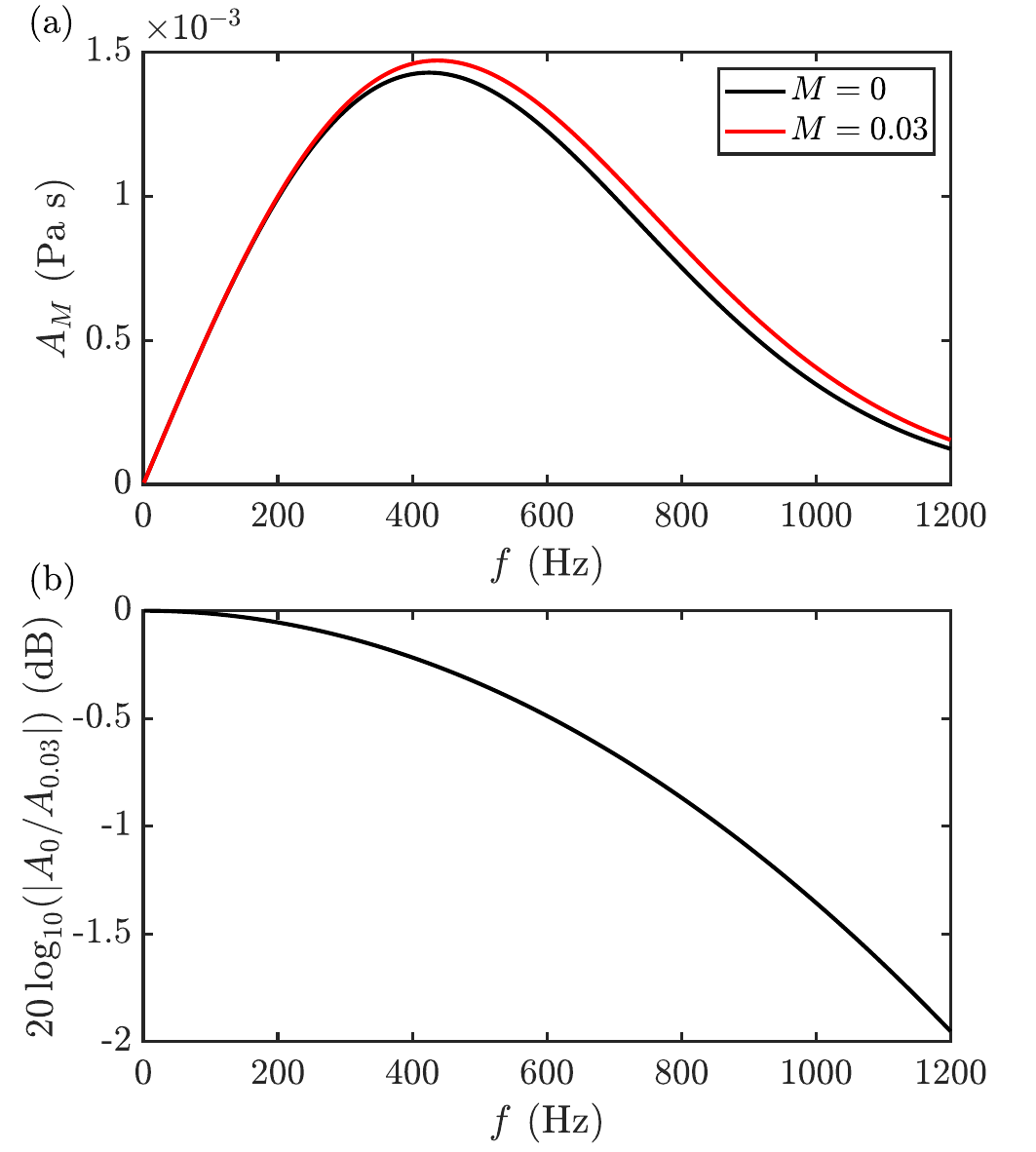}  
	\caption{(a) Amplitude of the Gaussian impulsive source without mean flow and with a mean flow of Mach number equal to 0.03 and (b) difference in dB.}
		\label{fig:amplitude}
\end{figure}

The amplitude of the Gaussian impulsive source thus depends on the surrounding mean velocity. As an illustration, Figure~\ref{fig:amplitude} shows the source amplitude $ A_M$ for $M=0$ and for $M=0.03$, which corresponds to the Mach number at the height of the wind turbine hub. The source amplitude tends to be shifted towards higher frequencies. The difference in dB is negligible at low frequencies. However, it becomes noticeable at high frequencies, even for this small Mach number. Thus, the difference  reaches around 1.5~dB at 1~kHz.

\subsubsection*{Evaluation of noise levels relative to the free field}

The calculation of the noise levels relative to the free field from the LEE  has to account for the  modification of the source amplitude by the presence of the mean flow. For that, we estimate the free-field solution as:
\begin{equation} \label{p_ff.eq}
\hat{p}_{\rm ff}(\mathbf{x},\omega) = -\dfrac{{\rm i}}{4}A_M(\omega)H_0^{(1)}(k_0 R) 
\end{equation}
where the amplitude of the source is determined with Eq.~(\ref{AM_def.eq}) considering the Mach number  at the source height instead of $M=0$. Eq.~(\ref{p_ff.eq}) is equivalent to Eq.~(\ref{eq:pff_LEE}).



\begin{thebibliography}{38}
  \def\enquote#1{``#1,''}
  \def\plainquote#1{``#1''}
  \expandafter\ifx\csname natexlab\endcsname\relax\def\natexlab#1{#1}\fi
  \providecommand{\dourl}[1]{\href{http://#1}{\nolinkurl{#1}}}
  \providecommand{\bibinfo}[2]{#2}
  \providecommand{\noopsort}[1]{}
  \providecommand{\switchargs}[2]{#2#1}
    \def\eatspace #1{#1}
  \providecommand{\dodoi}[1]{doi: \href{http://dx.doi.org/#1}{\nolinkurl{#1}}}
  
  \bibitem[{Attenborough \emph{et~al.}(2011)Attenborough, Bashir, and
    Taherzadeh}]{attenborough_outdoor_2011}
  \bibinfo{author}{Attenborough, K.}, \bibinfo{author}{Bashir, I.},  and
    \bibinfo{author}{Taherzadeh, S.} (\textbf{\bibinfo{year}{2011}}).
    \enquote{\bibinfo{title}{Outdoor ground impedance models}}
    \bibinfo{journal}{The Journal of the Acoustical Society of America}
    \textbf{129}(5), \bibinfo{pages}{2806--2819}, \dodoi{10.1121/1.3569740}.
  
  \bibitem[{Barlas \emph{et~al.}(2018)Barlas, Wu, Zhu, Porté-Agel, and
    Shen}]{barlas_variability_2018}
  \bibinfo{author}{Barlas, E.}, \bibinfo{author}{Wu, K.~L.},
    \bibinfo{author}{Zhu, W.~J.}, \bibinfo{author}{Porté-Agel, F.},  and
    \bibinfo{author}{Shen, W.~Z.} (\textbf{\bibinfo{year}{2018}}).
    \enquote{\bibinfo{title}{Variability of wind turbine noise over a diurnal
    cycle}} \bibinfo{journal}{Renewable Energy} \textbf{126},
    \bibinfo{pages}{791--800}, \dodoi{10.1016/j.renene.2018.03.086}.
  
  \bibitem[{Barlas \emph{et~al.}(2017{\natexlab{a}})Barlas, Zhu, Shen, Dag, and
    Moriarty}]{barlas_consistent_2017}
  \bibinfo{author}{Barlas, E.}, \bibinfo{author}{Zhu, W.~J.},
    \bibinfo{author}{Shen, W.~Z.}, \bibinfo{author}{Dag, K.~O.},  and
    \bibinfo{author}{Moriarty, P.} (\textbf{\bibinfo{year}{2017}}{\natexlab{a}}).
    \enquote{\bibinfo{title}{Consistent modelling of wind turbine noise
    propagation from source to receiver}} \bibinfo{journal}{The Journal of the
    Acoustical Society of America} \textbf{142}(5), \bibinfo{pages}{3297--3310},
    \dourl{http://asa.scitation.org/doi/10.1121/1.5012747},
    \dodoi{10.1121/1.5012747}.
  
  \bibitem[{Barlas \emph{et~al.}(2017{\natexlab{b}})Barlas, Zhu, Shen, Kelly, and
    Andersen}]{barlas_effects_2017}
  \bibinfo{author}{Barlas, E.}, \bibinfo{author}{Zhu, W.~J.},
    \bibinfo{author}{Shen, W.~Z.}, \bibinfo{author}{Kelly, M.},  and
    \bibinfo{author}{Andersen, S.~J.}
    (\textbf{\bibinfo{year}{2017}}{\natexlab{b}}).
    \enquote{\bibinfo{title}{Effects of wind turbine wake on atmospheric sound
    propagation}} \bibinfo{journal}{Applied Acoustics} \textbf{122},
    \bibinfo{pages}{51--61}, \dodoi{10.1016/j.apacoust.2017.02.010}.
  
  \bibitem[{Berland \emph{et~al.}(2006)Berland, Bogey, and
    Bailly}]{berland_low-dissipation_2006}
  \bibinfo{author}{Berland, J.}, \bibinfo{author}{Bogey, C.},  and
    \bibinfo{author}{Bailly, C.} (\textbf{\bibinfo{year}{2006}}).
    \enquote{\bibinfo{title}{Low-dissipation and low-dispersion fourth-order
    {Runge}–{Kutta} algorithm}} \bibinfo{journal}{Computers \& Fluids}
    \textbf{35}(10), \bibinfo{pages}{1459--1463},
    \dodoi{10.1016/j.compfluid.2005.04.003}.
  
  \bibitem[{Berland \emph{et~al.}(2007)Berland, Bogey, Marsden, and
    Bailly}]{berland_high-order_2007}
  \bibinfo{author}{Berland, J.}, \bibinfo{author}{Bogey, C.},
    \bibinfo{author}{Marsden, O.},  and \bibinfo{author}{Bailly, C.}
    (\textbf{\bibinfo{year}{2007}}). \enquote{\bibinfo{title}{High-order, low
    dispersive and low dissipative explicit schemes for multiple-scale and
    boundary problems}} \bibinfo{journal}{Journal of Computational Physics}
    \textbf{224}(2), \bibinfo{pages}{637--662},
    \dodoi{10.1016/j.jcp.2006.10.017}.
  
  \bibitem[{Blumrich and Heimann(2002)}]{blumrich_linearized_2002}
  \bibinfo{author}{Blumrich, R.},  and \bibinfo{author}{Heimann, D.}
    (\textbf{\bibinfo{year}{2002}}). \enquote{\bibinfo{title}{A linearized
    {Eulerian} sound propagation model for studies of complex meteorological
    effects}} \bibinfo{journal}{The Journal of the Acoustical Society of America}
    \textbf{112}(2), \bibinfo{pages}{446--455}, \dodoi{10.1121/1.1485971}.
  
  \bibitem[{Bogey and Bailly(2004)}]{bogey_family_2004}
  \bibinfo{author}{Bogey, C.},  and \bibinfo{author}{Bailly, C.}
    (\textbf{\bibinfo{year}{2004}}). \enquote{\bibinfo{title}{A family of low
    dispersive and low dissipative explicit schemes for flow and noise
    computations}} \bibinfo{journal}{Journal of Computational Physics}
    \textbf{194}(1), \bibinfo{pages}{194--214},
    \dodoi{10.1016/j.jcp.2003.09.003}.
  
  \bibitem[{Collino(1997)}]{collino_perfectly_1997}
  \bibinfo{author}{Collino, F.} (\textbf{\bibinfo{year}{1997}}).
    \enquote{\bibinfo{title}{Perfectly matched absorbing layers for the paraxial
    equations}} \bibinfo{journal}{Journal of Computational Physics}
    \textbf{131}(1), \bibinfo{pages}{164--180}, \dodoi{10.1006/jcph.1996.5594}.
  
  \bibitem[{Cosnefroy(2019)}]{cosnefroy_simulation_nodate}
  \bibinfo{author}{Cosnefroy, M.} (\textbf{\bibinfo{year}{2019}}).
    \enquote{\bibinfo{title}{Simulation numérique de la propagation dans
    l’atmosphère de sons impulsionnels et confrontations expérimentales.
    {Numerical} simulation of atmospheric propagation of impulse sound and
    experimental comparisons}} Ph.D. thesis, \bibinfo{school}{Ecole Centrale de
    Lyon}, \bibinfo{note}{2019LYSEC014}.
  
  \bibitem[{Cotté(2019)}]{cotte_extended_2019}
  \bibinfo{author}{Cotté, B.} (\textbf{\bibinfo{year}{2019}}).
    \enquote{\bibinfo{title}{Extended source models for wind turbine noise
    propagation}} \bibinfo{journal}{The Journal of the Acoustical Society of
    America} \textbf{145}(3), \bibinfo{pages}{1363--1371},
    \dodoi{10.1121/1.5093307}.
  
  \bibitem[{Dallois \emph{et~al.}(2001)Dallois, Blanc-Benon, and
    Juvé}]{dallois_wide-angle_2001}
  \bibinfo{author}{Dallois, L.}, \bibinfo{author}{Blanc-Benon, P.},  and
    \bibinfo{author}{Juvé, D.} (\textbf{\bibinfo{year}{2001}}).
    \enquote{\bibinfo{title}{A wide-angle parabolic equation for acoustic waves
    in inhomogebeous moving media: applications to atmospheric sound
    propagation}} \bibinfo{journal}{Journal of Computational Acoustics}
    \textbf{09}(02), \bibinfo{pages}{477--494},
    \dodoi{10.1142/S0218396X01000772}.
  
  \bibitem[{Dragna and Blanc-Benon(2014)}]{dragna_towards_2014}
  \bibinfo{author}{Dragna, D.},  and \bibinfo{author}{Blanc-Benon, P.}
    (\textbf{\bibinfo{year}{2014}}). \enquote{\bibinfo{title}{Towards realistic
    simulations of sound radiation by moving sources in outdoor environments}}
    \bibinfo{journal}{International Journal of Aeroacoustics} \textbf{13}(5-6),
    \bibinfo{pages}{405--426}, \dodoi{10.1260/1475-472X.13.5-6.405}.
  
    \bibitem[{Dragna \emph{et~al.}(2011)Dragna, Cott{\'e}, {Blanc-Benon}, and
    Poisson}]{dragnaTimeDomainSimulationsOutdoor2011b}
  \bibinfo{author}{Dragna, D.}, \bibinfo{author}{Cott{\'e}, B.},
    \bibinfo{author}{{Blanc-Benon}, P.},  and \bibinfo{author}{Poisson, F.}
    (\textbf{\bibinfo{year}{2011}}). \enquote{\bibinfo{title}{Time-{{Domain
    Simulations}} of {{Outdoor Sound Propagation}} with {{Suitable Impedance
    Boundary Conditions}}}} \bibinfo{journal}{AIAA Journal} \textbf{49}(7),
    \bibinfo{pages}{1420--1428}, \dodoi{10.2514/1.J050636}.


  \bibitem[{Dumortier \emph{et~al.}(2015)Dumortier, Vincent, and
    Deaconu}]{dumortier_acoustic_2015}
  \bibinfo{author}{Dumortier, B.}, \bibinfo{author}{Vincent, E.},  and
    \bibinfo{author}{Deaconu, M.} (\textbf{\bibinfo{year}{2015}}).
    \enquote{\bibinfo{title}{Acoustic control of wind farms}}
    \bibinfo{journal}{The European Wind Energy Association Conference{,} Paris{,}
    France{,} November 17-20,} \bibinfo{pages}{1--8}, \dodoi{hal-01233730}.
  
  \bibitem[{Gadde and Stevens(2021)}]{gadde_interaction_2021}
  \bibinfo{author}{Gadde, S.~N.},  and \bibinfo{author}{Stevens, R. J. A.~M.}
    (\textbf{\bibinfo{year}{2021}}). \enquote{\bibinfo{title}{Interaction between
    low-level jets and wind farms in a stable atmospheric boundary layer}}
    \bibinfo{journal}{Physical Review Fluids} \textbf{6}(1),
    \bibinfo{pages}{014603}, \dodoi{10.1103/PhysRevFluids.6.014603}.
  
  \bibitem[{Gadde \emph{et~al.}(2021)Gadde, Stieren, and
    Stevens}]{gadde_large-eddy_2021}
  \bibinfo{author}{Gadde, S.~N.}, \bibinfo{author}{Stieren, A.},  and
    \bibinfo{author}{Stevens, R. J. A.~M.} (\textbf{\bibinfo{year}{2021}}).
    \enquote{\bibinfo{title}{Large-eddy simulations of stratified atmospheric
    boundary layers: comparison of different subgrid models}}
    \bibinfo{journal}{Boundary-Layer Meteorology} \textbf{178}(3),
    \bibinfo{pages}{363--382}, \dodoi{10.1007/s10546-020-00570-5}.
  
  \bibitem[{Gal-Chen and Somerville(1975)}]{gal-chen_use_1975}
  \bibinfo{author}{Gal-Chen, T.},  and \bibinfo{author}{Somerville, R.~C.}
    (\textbf{\bibinfo{year}{1975}}). \enquote{\bibinfo{title}{On the use of a
    coordinate transformation for the solution of the {Navier}-{Stokes}
    equations}} \bibinfo{journal}{Journal of Computational Physics}
    \textbf{17}(2), \bibinfo{pages}{209--228},
    \dodoi{10.1016/0021-9991(75)90037-6}.
  
  \bibitem[{Gilbert and White(1989)}]{gilbert_application_1989}
  \bibinfo{author}{Gilbert, K.~E.},  and \bibinfo{author}{White, M.~J.}
    (\textbf{\bibinfo{year}{1989}}). \enquote{\bibinfo{title}{Application of the
    parabolic equation to sound propagation in a refracting atmosphere}}
    \bibinfo{journal}{The Journal of the Acoustical Society of America}
    \textbf{85}(2), \bibinfo{pages}{630--637}, \dodoi{10.1121/1.397587}.
  
  \bibitem[{Hansen \emph{et~al.}(2019)Hansen, Nguyen, Zajamšek, Catcheside, and
    Hansen}]{hansen_prevalence_2019}
  \bibinfo{author}{Hansen, K.~L.}, \bibinfo{author}{Nguyen, P.},
    \bibinfo{author}{Zajamšek, B.}, \bibinfo{author}{Catcheside, P.},  and
    \bibinfo{author}{Hansen, C.~H.} (\textbf{\bibinfo{year}{2019}}).
    \enquote{\bibinfo{title}{Prevalence of wind farm amplitude modulation at
    long-range residential locations}} \bibinfo{journal}{Journal of Sound and
    Vibration} \textbf{455}, \bibinfo{pages}{136--149},
    \dodoi{10.1016/j.jsv.2019.05.008}.
  
  \bibitem[{Heimann and Englberger(2018)}]{heimann_3d-simulation_2018}
  \bibinfo{author}{Heimann, D.},  and \bibinfo{author}{Englberger, A.}
    (\textbf{\bibinfo{year}{2018}}). \enquote{\bibinfo{title}{{3D}-simulation of
    sound propagation through the wake of a wind turbine: {Impact} of the diurnal
    variability}} \bibinfo{journal}{Applied Acoustics} \textbf{141},
    \bibinfo{pages}{393--402}, \dodoi{10.1016/j.apacoust.2018.06.005}.
  
    \bibitem[{Heimann \emph{et~al.}(2018)Heimann, Englberger, and
  Schady}]{heimannSoundPropagationWake2018}
\bibinfo{author}{Heimann, D.}, \bibinfo{author}{Englberger, A.},  and
  \bibinfo{author}{Schady, A.} (\textbf{\bibinfo{year}{2018}}).
  \enquote{\bibinfo{title}{Sound propagation through the wake flow of a hilltop
  wind turbine-{A} numerical study}} \bibinfo{journal}{Wind Energy}
  \textbf{21}(8), \bibinfo{pages}{650--662},
  \dodoi{10.1002/we.2185}.

  \bibitem[{Kayser \emph{et~al.}(2020)Kayser, Cotté, Ecotière, and
    Gauvreau}]{kayser_environmental_2020}
  \bibinfo{author}{Kayser, B.}, \bibinfo{author}{Cotté, B.},
    \bibinfo{author}{Ecotière, D.},  and \bibinfo{author}{Gauvreau, B.}
    (\textbf{\bibinfo{year}{2020}}). \enquote{\bibinfo{title}{Environmental
    parameters sensitivity analysis for the modeling of wind turbine noise in
    downwind conditions}} \bibinfo{journal}{The Journal of the Acoustical Society
    of America} \textbf{148}(6), \bibinfo{pages}{3623--3632},
    \dodoi{10.1121/10.0002872}.
  
  \bibitem[{Kayser \emph{et~al.}(2023)Kayser, Mascarenhas, Cotté, Ecotière, and
    Gauvreau}]{kayserValidityEffectiveSound2023}
  \bibinfo{author}{Kayser, B.}, \bibinfo{author}{Mascarenhas, D.},
    \bibinfo{author}{Cotté, B.}, \bibinfo{author}{Ecotière, D.},  and
    \bibinfo{author}{Gauvreau, B.} (\textbf{\bibinfo{year}{2023}}).
    \enquote{\bibinfo{title}{Validity of the effective sound speed approximation
    in parabolic equation models for wind turbine noise propagation}}
    \bibinfo{journal}{The Journal of the Acoustical Society of America}
    \textbf{153}(3), \bibinfo{pages}{1846--1854},
    \dodoi{10.1121/10.0017653}.
  
  \bibitem[{Komatitsch and Martin(2007)}]{komatitsch_unsplit_2007}
  \bibinfo{author}{Komatitsch, D.},  and \bibinfo{author}{Martin, R.}
    (\textbf{\bibinfo{year}{2007}}). \enquote{\bibinfo{title}{An unsplit
    convolutional {Perfectly} {Matched} {Layer} improved at grazing incidence for
    the seismic wave equation}} \bibinfo{journal}{Geophysical Journal
    International} \textbf{72}(1), \bibinfo{pages}{333--344},
    \dodoi{10.1190/1.2757586}.
  
  \bibitem[{Lee \emph{et~al.}(2016)Lee, Lee, and Honhoff}]{lee_prediction_2016}
  \bibinfo{author}{Lee, S.}, \bibinfo{author}{Lee, D.},  and
    \bibinfo{author}{Honhoff, S.} (\textbf{\bibinfo{year}{2016}}).
    \enquote{\bibinfo{title}{Prediction of far-field wind turbine noise
    propagation with parabolic equation}} \bibinfo{journal}{The Journal of the
    Acoustical Society of America} \textbf{140}(2), \bibinfo{pages}{767--778},
    \dodoi{10.1121/1.4958996}.
  
  \bibitem[{Liu and Stevens(2020)}]{liu_effects_2020}
  \bibinfo{author}{Liu, L.},  and \bibinfo{author}{Stevens, R. J. A.~M.}
    (\textbf{\bibinfo{year}{2020}}). \enquote{\bibinfo{title}{Effects of
    two-dimensional steep hills on the performance of wind turbines and wind
    farms}} \bibinfo{journal}{Boundary-Layer Meteorology} \textbf{176}(2),
    \bibinfo{pages}{251--269}, \dodoi{10.1007/s10546-020-00522-z}.
  
  \bibitem[{Nyborg \emph{et~al.}(2022)Nyborg, Fischer, Thysell, Feng,
    Søndergaard, Hansen, Hansen, and Bertagnolio}]{nyborg_propagation_2022}
  \bibinfo{author}{Nyborg, C.~M.}, \bibinfo{author}{Fischer, A.},
    \bibinfo{author}{Thysell, E.}, \bibinfo{author}{Feng, J.},
    \bibinfo{author}{Søndergaard, L.~S.}, \bibinfo{author}{Hansen, T.~R.},
    \bibinfo{author}{Hansen, K.~S.},  and \bibinfo{author}{Bertagnolio, F.}
    (\textbf{\bibinfo{year}{2022}}). \enquote{\bibinfo{title}{Propagation of wind
    turbine noise: measurements and model evaluation}} \bibinfo{journal}{Journal
    of Physics} \bibinfo{pages}{14}, \dodoi{10.1088/1742-6596/2265/3/032041}.
  
  \bibitem[{Ostashev \emph{et~al.}(2005)Ostashev, Wilson, Liu, Aldridge, Symons,
    and Marlin}]{ostashev_equations_2005}
  \bibinfo{author}{Ostashev, V.~E.}, \bibinfo{author}{Wilson, D.~K.},
    \bibinfo{author}{Liu, L.}, \bibinfo{author}{Aldridge, D.~F.},
    \bibinfo{author}{Symons, N.~P.},  and \bibinfo{author}{Marlin, D.}
    (\textbf{\bibinfo{year}{2005}}). \enquote{\bibinfo{title}{Equations for
    finite-difference, time-domain simulation of sound propagation in moving
    inhomogeneous media and numerical implementation}} \bibinfo{journal}{The
    Journal of the Acoustical Society of America} \textbf{117}(2),
    \bibinfo{pages}{503--517}, \dodoi{10.1121/1.1841531}.
  
  \bibitem[{Ostashev \emph{et~al.}(2020)Ostashev, Wilson, and
    Muhlestein}]{ostashev_wave_2020}
  \bibinfo{author}{Ostashev, V.~E.}, \bibinfo{author}{Wilson, D.~K.},  and
    \bibinfo{author}{Muhlestein, M.~B.} (\textbf{\bibinfo{year}{2020}}).
    \enquote{\bibinfo{title}{Wave and extra-wide-angle parabolic equations for
    sound propagation in a moving atmosphere}} \bibinfo{journal}{The Journal of
    the Acoustical Society of America} \textbf{147}(6),
    \bibinfo{pages}{3969--3984}, \dodoi{10.1121/10.0001397}.
  
  \bibitem[{Prospathopoulos and
    Voutsinas(2007)}]{prospathopoulos_application_2007}
  \bibinfo{author}{Prospathopoulos, J.~M.},  and \bibinfo{author}{Voutsinas,
    S.~G.} (\textbf{\bibinfo{year}{2007}}). \enquote{\bibinfo{title}{Application
    of a ray theory model to the prediction of noise emissions from isolated wind
    turbines and wind parks}} \bibinfo{journal}{Wind Energy} \textbf{10}(2),
    \bibinfo{pages}{103--119}, \dodoi{10.1002/we.211}.
  
  \bibitem[{Rienstra(2006)}]{rienstra_impedance_2006}
  \bibinfo{author}{Rienstra, S.} (\textbf{\bibinfo{year}{2006}}).
    \enquote{\bibinfo{title}{Impedance models in time domain, including the
    extended {Helmoltz} resonator model}} \bibinfo{publisher}{12th {AIAA}/{CEAS}
    {Aeroacoustics} {Conference} (27th {AIAA} {Aeroacoustics} {Conference}),
    American Institute of Aeronautics and Astronautics},
    \dodoi{10.2514/6.2006-2686}.
  
  \bibitem[{Sack and West(1995)}]{sack_parabolic_1995}
  \bibinfo{author}{Sack, R.~A.},  and \bibinfo{author}{West, M.}
    (\textbf{\bibinfo{year}{1995}}). \enquote{\bibinfo{title}{A parabolic
    equation for sound propagation in two dimensions over any smooth terrain
    profile: {The} generalised terrain parabolic equation ({GT}-{PE})}}
    \bibinfo{journal}{Applied Acoustics} \textbf{45}(2),
    \bibinfo{pages}{113--129}, \dodoi{10.1016/0003-682X(94)00039-X}.
  
  \bibitem[{Salomons(2001)}]{salomons_computational_2001}
  \bibinfo{author}{Salomons, E.~M.} (\textbf{\bibinfo{year}{2001}}).
    \emph{\bibinfo{title}{Computational atmospheric acoustics}}
    (\bibinfo{publisher}{Springer Netherlands}, \bibinfo{address}{Dordrecht}).
  
  \bibitem[{Salomons \emph{et~al.}(2002)Salomons, Blumrich, and
    Heimann}]{salomons_eulerian_2002}
  \bibinfo{author}{Salomons, E.~M.}, \bibinfo{author}{Blumrich, R.},  and
    \bibinfo{author}{Heimann, D.} (\textbf{\bibinfo{year}{2002}}).
    \enquote{\bibinfo{title}{Eulerian time-domain model for sound propagation
    over a finite-impedance ground surface. comparison with frequency-domain
    models}} \bibinfo{journal}{Acta Acustica} \textbf{88}(4),
    \bibinfo{pages}{483–492}.
  
  \bibitem[{Sessarego and Shen(2020)}]{sessarego_noise_2020}
  \bibinfo{author}{Sessarego, M.},  and \bibinfo{author}{Shen, W.~Z.}
    (\textbf{\bibinfo{year}{2020}}). \enquote{\bibinfo{title}{Noise propagation
    calculation of a wind turbine in complex terrain}} \bibinfo{journal}{Journal
    of Physics: Conference Series} \textbf{1452}(1), \bibinfo{pages}{012063},
    \dodoi{10.1088/1742-6596/1452/1/012063}.
  
  \bibitem[{Shen \emph{et~al.}(2019)Shen, Zhu, Barlas, and
    Li}]{shen_advanced_2019}
  \bibinfo{author}{Shen, W.~Z.}, \bibinfo{author}{Zhu, W.~J.},
    \bibinfo{author}{Barlas, E.},  and \bibinfo{author}{Li, Y.}
    (\textbf{\bibinfo{year}{2019}}). \enquote{\bibinfo{title}{Advanced flow and
    noise simulation method for wind farm assessment in complex terrain}}
    \bibinfo{journal}{Renewable Energy} \textbf{143},
    \bibinfo{pages}{1812--1825}, \dodoi{10.1016/j.renene.2019.05.140}.
  
  \bibitem[{Tian and Cotté(2016)}]{tian_wind_2016}
  \bibinfo{author}{Tian, Y.},  and \bibinfo{author}{Cotté, B.}
    (\textbf{\bibinfo{year}{2016}}). \enquote{\bibinfo{title}{Wind turbine noise
    modeling based on amiet's theory: effects of wind shear and atmospheric
    turbulence}} \bibinfo{journal}{Acta Acustica united with Acustica}
    \textbf{102}(4), \bibinfo{pages}{626--639}, \dodoi{10.3813/AAA.918979}.
  
  \bibitem[{Troian \emph{et~al.}(2017)Troian, Dragna, Bailly, and
    Galland}]{troian_broadband_2017}
  \bibinfo{author}{Troian, R.}, \bibinfo{author}{Dragna, D.},
    \bibinfo{author}{Bailly, C.},  and \bibinfo{author}{Galland, M.-A.}
    (\textbf{\bibinfo{year}{2017}}). \enquote{\bibinfo{title}{Broadband liner
    impedance eduction for multimodal acoustic propagation in the presence of a
    mean flow}} \bibinfo{journal}{Journal of Sound and Vibration} \textbf{392},
    \bibinfo{pages}{200--216}, \dodoi{10.1016/j.jsv.2016.10.014}.
  
  \bibitem[{Van~Renterghem(2014)}]{van_renterghem_efficient_2014}
  \bibinfo{author}{Van~Renterghem, T.} (\textbf{\bibinfo{year}{2014}}).
    \enquote{\bibinfo{title}{Efficient outdoor sound propagation modeling with
    the finite-difference time-domain fdtd method: a review}}
    \bibinfo{journal}{International Journal of Aeroacoustics} \textbf{13}(5-6),
    \bibinfo{pages}{385--404}, \dodoi{10.1260/1475-472X.13.5-6.385}.
  
  \end{thebibliography}


\end{document}